\documentclass[preprint]{aastex}
\usepackage{longtable}
\usepackage{color}
\everymath{\rm}
\everydisplay{\sf}
\begin{document}
\title{Alpha Element Abundances in a Large Sample of Galactic Planetary Nebulae}
\shorttitle{$\alpha$ Element Abundances in Galactic PNe}
\shortauthors{Milingo et al.}
\author{J.B. Milingo\altaffilmark{1,2}}
\affil{Department of Physics, Gettysburg College, Gettysburg, PA 17325}
\email{jmilingo@gettysburg.edu}

\author{K.B. Kwitter\altaffilmark{2}}
\affil{Department of Astronomy, Williams College, Williamstown, MA 01267}
\email{kkwitter@williams.edu}

\author{R.B.C. Henry\altaffilmark{2}}
\affil{Department of Physics $\&$ Astronomy, University of Oklahoma, Norman, OK 73019}
\email{henry@mail.nhn.ou.edu}

\and

\author{S.P. Souza\altaffilmark{1,2}}
\affil{Department of Astronomy, Williams College, Williamstown, MA 01267}
\email{ssouza@williams.edu}

\altaffiltext{1}{Visiting Astronomer, Cerro Tololo Inter-American Observatory, National Optical Astronomy Observatory, which is operated by the Association of Universities for Research in Astronomy (AURA), Inc., under cooperative agreement with the National Science Foundation.}
\altaffiltext{2}{Visiting Astronomer, Kitt Peak National Observatory, National Optical Astronomy Observatory, which is operated by the Association of Universities for  Research in Astronomy (AURA), Inc., under cooperative agreement with the National Science Foundation.}

\begin{abstract}
In this paper we present emission line strengths, abundances, and element ratios (X/O for Ne, S, Cl, and Ar) for a sample of 38 Galactic disk planetary nebulae (PNe) consisting primarily of Peimbert classification Type I.   Spectrophotometry for these PNe incorporates an extended optical/near-IR range of $\lambda\lambda$3600-9600 $\AA$ including the [\ion{S}{3}] lines at 9069$\AA$ and 9532$\AA$, setting this relatively large sample apart from typical spectral coverage.  We've utilized ELSA (Emission Line Spectrum Analyzer), a 5-level atom abundance routine, to determine ${T_e}, {N_e}$, ICFs, and total element abundances, thereby continuing our work toward a uniformly processed set of data.  With a compilation of data from $>$120 Milky Way planetary nebulae (PNe), we present results from our most recent analysis of abundance patterns in Galactic disk PNe.  With a wide range of metallicities, galactocentric distances, and both Type I and non-Type I objects, we've examined the alpha elements against HII regions and blue compact galaxies (H2BCG) to discern signatures of depletion or enhancement in PNe progenitor stars, particularly the destruction or production of O and Ne. We present evidence that many PNe have higher Ne/O and lower Ar/Ne ratios compared to H2BCGs within the range of 8.5-9.0 for 12+log(O/H). This suggests that Ne is being synthesized in the low-intermediate mass progenitors.  Sulfur abundances in PNe continue to show great scatter and are systematically lower than those found in H2BCG at a given metallicity.  Although we find that PNe do show some distinction in alpha elements when compared to H2BCG, within the Peimbert classification types studied PNe do not show significant differences in alpha elements amongst themselves, at least to an extent that would distinguish in situ nucleosynthesis from the observed dispersion in abundance ratios. 
\end{abstract}


\section{Introduction}
Planetary Nebulae (PN) abundance patterns have long been used to note signatures of nuclear processing and to trace the distribution of metals throughout galaxies.  Being the shed envelopes resulting from the late evolutionary stages of intermediate-mass stars (1$M_\sun \leq$ M $\leq 8M_\sun$), PNe abundances reflect both the evolution of their progenitors and the natal interstellar medium from which they formed.  PNe are populous, cover a range in progenitor mass and metallicity, and are distributed throughout the disk, bulge, and halo of our Galaxy; hence their abundances are useful probes into the evolution of intermediate mass stars, the yields of previous generations of stars, and the chemical evolution of galaxies.   

This work continues our compilation of abundances in Galactic PNe that are based upon newly acquired CCD spectrophotometry over an extended range of optical/near IR wavelengths including the strong [S III] emission features at $\lambda$9069 and $\lambda$9532 \AA.  Here we present new spectrophotometric data for 38 mostly Type I Galactic PNe. These new spectra, when added to our previous samples \citep{IIB, three, HKB04} bring the total sample to $>$120, covering a substantial range in galactocentric distance and including Peimbert classification Types I and II as well as several anticenter PNe.   Peimbert "types" are categorized by the degree to which N and He have been enhanced \citep{Peimbert1978}.  This is indicative of progenitor mass insofar as there is a mass dependence on the sites of nucleosynthesis that create and alter the abundances we measure in PNe.  Type I PNe have high N and He which is consistent with them having progenitors $\gtrsim$ 2.5 $M\sun$, whereas Type II PNe are inferred to have less massive progenitors \citep{Peimbert1978, Maciel92, KB94}.  With this data in hand we are compelled to look for distinguishing signatures of nuclear self-contamination across the range of metallicity, Peimbert "type", and galactocentric distance our sample contains.  The questions of oxygen and neon depletion and/or enhancement as seen in the PN gas are of particular interest.  

The evolution of PNe progenitors gives rise to convective episodes that reach down into temperature regions where nuclear processing can occur.  The subsequent dredge-up of products from this in situ nucleosynthesis alters the natal envelope material.  The first dredge-up (FDU) occurs on the first giant branch and is a significant mixing event for the lower-mass end of PNe progenitors $<$ 1.5 $M\sun$ \citep{BandS99}.  While on the asymptotic giant branch (AGB) the third dredge-up (TDU) and hot bottom burning (HBB) may occur, depending on the mass and metallicity of the progenitor.  The ejected envelopes of these evolved stars become planetary nebulae, so any surface modification in the chemical composition of the progenitor star should be seen in its PN gas \citep{vG97, RV81}.  It is important to note here that Type I PNe have long been connected to bipolar morphology \citep{PTP83} and that bipolar morphology in PNe is most likely the result of binary progenitors \citep{dM09}.  The effects of binary interactions on the evolution of the progenitor and the composition of the subsequent PNe gas could be at play here but are beyond the scope of this paper.

Within the published literature there is a lack of consistent observational support for the production/depletion of oxygen and neon in PN abundances. Oxygen has a significant history here as it is a common metallicity tracer for the interstellar medium; marking the degree to which chemical enrichment has occurred.  However oxygen is problematic in PNe as it is suspected of self-contamination, in other words undergoing nuclear processing during the evolution of the progenitor star.  Oxygen is subject to depletion through the ON cycle during HBB \citep{Clayton68}, and possibly enrichment via the TDU of $^{16}O$ from the products of He burning and subsequent alpha capture \citep{IR83, Pequignot2000}.  The degree to which the surface abundance of oxygen is expected to be altered is unclear as it is subject to the parameter space of dredge-up episodes and HBB \citep{RV81, Charbonnel2005}.  In PNe progenitors $\sim$ 3-5 $M\sun$, oxygen depletion owing to HBB could be modestly visible though difficult to discern within the typical uncertainties associated with PN abundance work (0.1-0.3 dex).  Alternate tracers such as sulfur and argon, which are believed to be precluded from self-contamination, could instead be used to indicate the metallicity of the natal ISM of PN progenitors.  Neon also has a history as a metallicity tracer but it is also suspected of undergoing self-contamination.  Neon can be enriched through the production of $^{22}Ne$ in TP-AGB stars via two ${\alpha}$ captures onto $^{14}N$ \citep{Busso2006, KL2003, Marigo2003}.  According to \citet{KL2003} the production and dredge-up of $^{22}Ne$ is particularly efficient in progenitor stars $\sim$ 2-4 $M\sun$ across a range of metallicities.  Oxygen, neon, sulfur, and argon are believed to be produced in ÒlockstepÓ, or proportionally to each other, in massive stars \citep{Henry1989, HKB04}.  If we do expect elemental abundances to be altered due to self-contamination and subsequent dredge-up, then observing patterns that indicate depletion or enhancement across the parameter space of PNe serves as observational support of our understanding of the nuclear processing and convective episodes that occur during the late stages of low and intermediate-mass stellar evolution. 

Observationally it has proven difficult to note modest depletions and/or enhancements in alpha elements in PN.  The scatter seen in observed PN abundances is due to many factors including uncertainties in total element abundances, possible systematic effects brought on by compiling inhomogeneous data sets, as well as the true spread in elemental abundances that is expected due to the range of PN progenitor masses, metallicities, galactocentric distance (abundance gradient effects), and the degree of nuclear self-contamination.  We are attempting to address some of this using extended spectral coverage, allowing for good quality sulfur abundances, and a uniformly processed set of data, in other words internally consistent observation, reduction, and abundance determination techniques.  

Subsequent sections of this paper discuss data acquisition and reduction, our abundance determination scheme, as well as results and analysis of X/H and X/O ratios for O, Ne, S, Cl, Ar, and N.  An important feature of our work is the use of the near-IR [\ion{S}{3}] lines and, where available, a [\ion{S}{3}] temperature to calculate $S^{+2}$ abundances.  Because a significant portion of S in PNe resides in $S^{+2}$ and higher ionization stages, including the near-IR [\ion{S}{3}] lines improves the extrapolation from observed ion abundances to total element abundance. In $\S$ 4.2 we discuss issues related to observed sulfur abundances in PNe.  

\section{Observations and Reductions}

We've acquired spectra for 38 Galactic disk PNe with coverage that extends from 3600\AA\ to 9600\AA.  Using the same instrumentation and setup as our previous Type II PNe program \citep{IIA}, observations were obtained over a total of four runs; Kitt Peak National Observatory (KPNO) in June 2003 $\&$ August 2004, and Cerro Tololo Interamerican Observatory (CTIO) in November 2003 $\&$ August 2004.  Table 1 lists the objects, total exposure times in the blue and red grating configurations, and the run during which the spectra were acquired.  

To obtain the needed range of spectral coverage, at both observatories we used a combination of two gratings with overlap in the region of $H\alpha$ for subsequent merging. The KPNO observations utilized the 2.1m telescope and GoldCam spectrograph.  We used a long E-W slit with projected dimensions of 5\arcsec x 305\arcsec  to accommodate angularly large objects and simultaneous "sky" exposure.  Spatial scale for the GoldCam is 0.78 \arcsec/pixel.   On the GoldCam we used gratings 240 $\&$ 58 in first order with nominal wavelength dispersions of 1.5 \AA/pixel and 1.9 \AA/pixel respectively.  This gave us a FWHM resolution of ~8\AA\ and 10\AA\  respectively.  

The CTIO observations were made with the 1.5m telescope and the R-C (Cassegrain) spectrograph using a long E-W oriented slit with projected dimensions of 5\arcsec x 321\arcsec.  The spatial scale for the R-C spectrograph is 1.3 \arcsec/pixel.  Gratings 9 $\&$ 22 were used in first order with nominal wavelength dispersion of 2.8 \AA/pixel and 8.6\AA\  FWHM resolution for both.      

The angular sizes within our Type I sample vary greatly, so we either centered the PNe in the slit or concentrated on a bright portion of the nebula.  Any offsets from the center of the PNe are listed in Table~1.  If possible the central star was avoided during the observations and extraction process, but for angularly small objects it is unavoidably included.  

Standard calibration frames were obtained for bias subtraction, flat-fielding, slit illumination correction, and wavelength calibration.  Standard star spectra were also obtained for flux calibration.  The removal of instrumental signature, calibration, and extraction of the final one-dimensional spectra were done using the {\it twodspec} package in IRAF\footnote{IRAF is distributed by the National Optical Astronomy Observatories (NOAO), which are operated by the Association of Universities for Research in Astronomy (AURA), Inc., under cooperative agreement with the National Science Foundation.}.  At both KPNO and CTIO, the thinned red chips produce interference fringes visible in the red. In our spectra the fringes appear at the $\sim$$\pm$1\% level at $\sim$7500\AA\ and increase in amplitude with increasing wavelength: $\pm$1-2\% at 8000\AA, $\pm$4-6\% at 8500\AA, $\pm$6-8\% at 9000\AA. Even at their worst, {\it i.e.\/}, at $\sim$$\lambda$9500, the longest wavelength we measure, the fringe amplitude reaches only
about $\pm$7-10\%, and we note this additional uncertainty in our line intensities longward of $\sim$7500\AA. 

\section{Line Strengths and Abundances}

Raw line fluxes were measured in IRAF using the {\it splot} task.  For each object the red and blue ends of the spectrum were merged using the overlap in the ${H\alpha}$ region.  Table 2 lists the raw and dereddened line strengths for our sample of PNe.  Columns labeled {\it F(${\lambda}$)} give raw uncorrected fluxes as measured in {\it splot}.  These {\it F(${\lambda}$)} values have been scaled to H${\beta}$ = 100 using the observed value of F{\it (H${\beta}$)} for each object.  Intensities of strong lines have measurement uncertainties of ${\leq}$10\%, single colons indicate uncertainties of ${\gtrsim}$25\%, and double colons indicate uncertainties ${\gtrsim}$50\%.   The last three rows of the table give the extinction measure, {\it c}, for each object, the final converged values for ${H\alpha/H\beta}$, and the log of the observed ${H\beta}$ flux in units of ${ergs~cm^{-2}~s^{-1}}$.  Final dereddened intensity ratios are calculated as follows:
\begin{displaymath}
\frac{I(\lambda)}{I(H\beta)}=\frac{F(\lambda)}{F(H\beta)}x10^{cf(\lambda)},
\end{displaymath}
where I($\lambda$)/I(H$\beta$) is the final intensity ratio of an emission line to H$\beta$, F$\lambda$/F(H$\beta$) is the observed flux ratio of the line to H$\beta$, {\it c} is the extinction measure, and f($\lambda$) is the reddening function.

To calculate ionic and total element abundances we have employed ELSA (Emission Line Spectrum Analyzer), described in detail by \citet{JLHK2006}. ELSA incorporates a 5-level atom abundance program, spectrum merging and de-reddening functions \citep{SM1979}, and routines for determining values for the interstellar reddening measure ({\it c}) and nebular temperature and density (${T_e}$ and ${N_e}$).  If there is a non-zero flux present for \ion{He}{2} $\lambda$4686, ELSA corrects for contamination of the Balmer lines due to ionized helium.  This iterative correction exploits the temperature and density dependence of the recombination coefficients for hydrogen and helium, ultimately converging on a value for the intrinsic ${H\alpha/H\beta}$ ratio. ELSA also allows for error propagation from measured line strength uncertainty; for example, the uncertainty in a final line intensity includes the estimated uncertainty in the line flux as well as the uncertainty in {\it c}.  Uncertainties in temperature and density diagnostics incorporate the errors in their constituent lines, and ionic abundance uncertainties include errors in the relevant line intensities, temperature, and density. Final abundance uncertainties incorporate errors in the ionic abundances.    ELSA does not consider uncertainty in the ionization correction factors (ICF), nor does it include any systematic sources of uncertainty (e.g., flux calibration errors).    

Table 3 lists the electron temperatures and densities calculated for each object.  The temperatures have been determined from diagnostic emission line ratios of [\ion{O}{3}], [\ion{N}{2}], [\ion{O}{2}], [\ion{S}{2}], and [\ion{S}{3}].  Electron densities were determined using emission line ratios of [\ion{S}{2}] and [\ion{Cl}{3}] (see the footnotes to Table 3 for more detail).   Uncertainties in $T_e$ and $N_e$ are propagated from the uncertainties in the line fluxes used to calculate them.  

In calculating ionic abundances, ELSA uses the temperature appropriate to the ionization stage whenever possible, leading to a two-zone model. The [\ion{O}{3}] temperature is used to calculate ionic abundances for $He^+, He^{+2}, O^{+2}, Ar^{+2}$ and higher, $Cl^{+2}$ and higher, $Ne^{+2}$ and higher, and $S^{+2}$ when ${T_{[S~III]}}$ is not available or differs by more than 5000K from ${T_{[O~III]}}$. The [\ion{N}{2}] temperature is used for all other ions:  $O^0, O^+, N^+, S^+, C^{+2}$ and $Cl^+$.  Table 4 gives the temperature used in determining the ionic abundance for each observed transition.  
 
The general scheme for determining total element abundances is to sum the observed ionic contributions and correct for any unseen ionization stages with an ionization correction factor (ICF).  We symbolically represent this scheme as:

\begin{displaymath}
N(X) = \bigg\{\sum_{\lambda}\frac{I_\lambda}{\epsilon_\lambda(T_e, N_e)}\bigg\}  \cdot ICF(X).
\end{displaymath}

Here ${I_\lambda}$ is the dereddened ${H\beta}$-scaled value of the line strengths of observed ionic species, the ${\epsilon_\lambda}$ factor is the energy generated per transition for a particular ion, ICF(X) is the ionization correction factor determined using the functional forms given in \citet{KH2001}, and finally N(X) is the total number abundance of element X with respect to H.  

Table 4 shows the observed ionic species and associated transition wavelengths (in ${\AA}$) in column 1, their respective number abundances (as determined by that particular transition) in column 3, and the temperatures used to determine those abundances in column 2.  [\ion{S}{2}] densities were used for all ionic abundance calculations.  Each final ionic abundance is given as the mean of the observed transitions (those marked with an asterisk), weighted by flux.  The ionization correction factors used in calculating the total element abundances are also shown.  As stated before ELSA does not consider uncertainty in the ICF.  From \citet{KH2001} the uncertainty in the ICF for oxygen is approximately 10\% and the ICFs for other elements are uncertain by 20-30\%.  Table 5 shows the total element abundances X/H and element ratios X/O for each object.   

\section{Analysis and Discussion}

\subsection{Alpha Elements}
The primary motivation for this project is to test for evidence of alpha element destruction or synthesis during the evolution of PN progenitors.  The abundances of Ne, O, S, Cl\footnote{While Cl is technically not an alpha element, since its stable isotopes Cl$^{35}$ and Cl$^{37}$ have odd atomic masses, we nevertheless include it in our study of alpha elements because it arises as a secondary product during oxygen burning in stars when alpha particles are in high concentration \citep{Clayton2003}.}, and Ar have been measured in nearly all of the sample objects. Thus it is possible to study their behavior relative to one another in PNe, as well as to compare abundance patterns in PNe with what is observed in a combined sample of H~II regions and blue compact galaxies (BCGs). The latter two object types provide probes of relatively homogeneous interstellar material which generally show little evidence of abundance enhancements produced by the evolved stars harbored within them \citep{W09}.  If these alpha elements are produced in lockstep from a given demographic distribution of stars, we should see their co-evolution clearly displayed in PNe and H2BCG.  If PNe progenitors are altering these alpha elements in the late stages of evolution it can potentially show up as a departure from the lockstep behavior.  

Figures~\ref{nevo}-\ref{clvar} show plots of the ten element pairs, where the quantity on each axis is the logarithm of the number abundance of the element with respect to H, assuming that logH=12. We include three Galactic PN samples, that of \citet{HKB04} of mostly Type~II PNe, the Type~I objects of the current paper, and PNe located in the anticenter direction from \citet{KK09}. These three samples are referred to in the legends as HKB04, Milingo09, and Kwitter09, respectively. In addition we have added LMC objects recently studied by \citet{BS08}, referred to in the legends as BS, for the discussion of sulfur abundances in PNe.  In all, the complete PN sample covers a metallicity range of roughly 7.6 to 8.9 in 12+log(O/H). For comparison, we note that the sample analyzed by \citet{WL08} extend down to 7.3.

The H~II baseline in the figures comprises data of blue compact galaxies from \citet{IT99}, Galactic H~II regions from \citet{S83} and \citet{E98}, H~II region data for external galaxies from \citet{S75}, and M101 data from \citet{KBG03}\footnote{We note that there is a paucity of neon abundance data in high metallicity H~II regions. This stems partially from the fact that in metal-rich systems, the [Ne~III] emissions are often weak, resulting in poor signal-to-noise ratios.}. Collectively these will be referred to henceforth as H2BCG. Finally, abundance data for the sun \citep{A05} appear in our plots as well. Symbols are defined in the figure legends. We note that Cl abundances were not available in the H2BCG data, and thus Figs.~\ref{clvo}, \ref{clvne}, \ref{clvs}, and \ref{clvar} show PN abundances only.  It should be noted that we are making an a priori distinction between Type I and non-Type I PNe.  As discussed in \citet{HKB04} Type I PNe have enhanced nitrogen (and often helium) abundances and are inferred to come from more massive progenitors hence from a more recent epoch.  Type II PNe have less N and He enhancement and are inferred to come from a more intermediate age population with lower progenitor masses than Type I PNe.  The original Type I classification included both enhanced He/H and N/O.  In this work we continue to follow the discussion of \citet{KB94} who use only nitrogen as a discriminant, setting the minimum nitrogen abundance for Type I PNe as the sum of the carbon and nitrogen abundances seen in the H II regions in the host galaxy.    Given this \citet{KB94} calculate the minimum N/O ratio for Type I PNe in the Galaxy to be 0.8.  Using more recent abundance measurements for the Orion Nebula \citep{E98} and for the Sun \citep{GS98, A05}.  \citet{HKB04} derive a lower minimum N/O ratio for Type I classification.  For this paper all PNe with N/O ${\geq}$ 0.65 are classified as Type I.  We first discuss the establishment of the H~II baseline used in the figures before proceeding with our consideration of the PN data.

\subsubsection{H~II Regions and Blue Compact Galaxies: A baseline for comparison}

In Figs.~\ref{nevo}-\ref{clvar} the most notable feature is the relatively tight linear behavior (in log-log space) demonstrated by the H2BCG objects (except in the cases of Cl, as noted). The parameters for the least squares fits to the H2BCG data, shown with solid lines in the figures, are provided in the top section of Table~\ref{lsf}, where the left column identifies the relation, while columns 2-5 give values of the y intercept, slope,  correlation coefficient, and number of sample objects included, respectively. Quantities in parentheses indicate the number of outliers removed before calculating the fit.  Note that uncertainties for individual sample objects have not been taken into account. 
The first six relations (the last two will be discussed later)  possess very high correlation coefficients, indicating the correlations are unlikely to be statistical accidents. 

Results for H2BCGs in Table~\ref{lsf} show that all slopes are consistent with a value of unity, in agreement with the conventional wisdom regarding alpha element production. This last point is of great interest, since over a slightly larger metallicity range (7.3-8.8) \citet{WL08} inferred a slope for Ne vs. O of +1.14$\pm$.01 for a sample of extragalactic H~II regions and concluded that the rate of cosmic Ne evolution increases with metallicity. However, we find that the average log(Ne/O) for objects with 12+log(O/H)$<$8.5 is -0.70 while the average for objects above this threshold is -0.75, i.e., within the uncertainties the two sets are indistinguishable regarding Ne/O. 

Finally, we employed the chemical evolution code developed by R.B.C. Henry [see \citet{HP07} for a detailed discussion of the code] to predict interstellar abundances of O, Ne, S, Cl, and Ar as a function of time over the age of the Galaxy.  We then formed theoretical element to element plots like those in Figs.~\ref{nevo}-\ref{clvar}, calculated slopes, and compared the slopes with their observed counterparts. The models included the massive star yields for O, Ne, and S from \citet{Port98} and Cl and Ar yields from \citet{WW95} and SNIa yields of \citet{Nomoto97}, while the yields of low and intermediate mass stars taken from \citet{M01} do not include these elements. Thus, we are only testing the influences of massive stars and SNIa on alpha element behavior. Table~\ref{models} provides a comparison of the model-predicted slopes with those in column~3 of Table~\ref{lsf}, where we see good consistency between observation and theory, especially in light of the large uncertainties usually associated with stellar yield predictions. However, because of the yield uncertainties, we are unable to use the models to test whether there is a small increase of Ne/O with metallicity, as suggested by \citet{WL08}.

We note that the sun's position in Fig.~\ref{nevo} falls below the H2BCG trend line by about 0.15-0.20 dex, according to the solar abundances of \citet{A05}. This offset is consistent with the recommendation of \citet{WL08} to raise the solar Ne value by 0.22~dex. Since we conclude in the next subsection that Ne is enhanced in PNe, we would not recommend raising the solar Ne value to an extent which would place it above the trend line.

In summary, the evidence found in H2BCG objects appears to be consistent with the expected lockstep behavior among the alpha elements. This conclusion is at odds with that of  \citet{WL08} in the case of Ne versus O.

\subsubsection{Planetary Nebulae}

\begin{figure}
  \includegraphics[width=4in,angle=0]{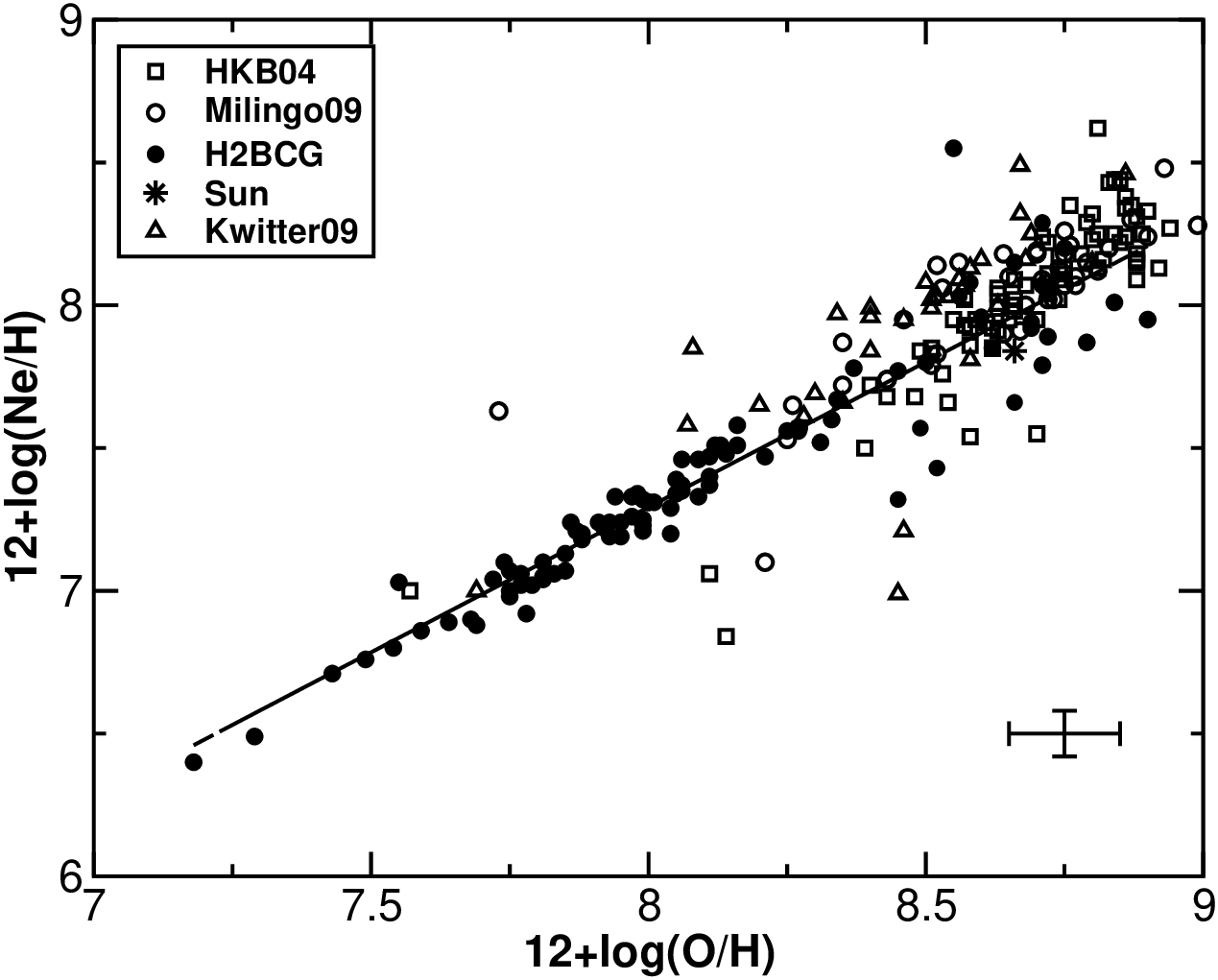} 
  \caption{12+log(Ne/H) versus 12+log(O/H) for PN data (HKB04, Milingo09, and Kwitter09) along with H~II region and blue compact galaxy abundance data (H2BCG) and the Sun.  References are provided in the text. The solid line is a least squares fit to the H2BCG data.}
  \label{nevo}
\end{figure}

\begin{figure}
   \includegraphics[width=3.9in,angle=0]{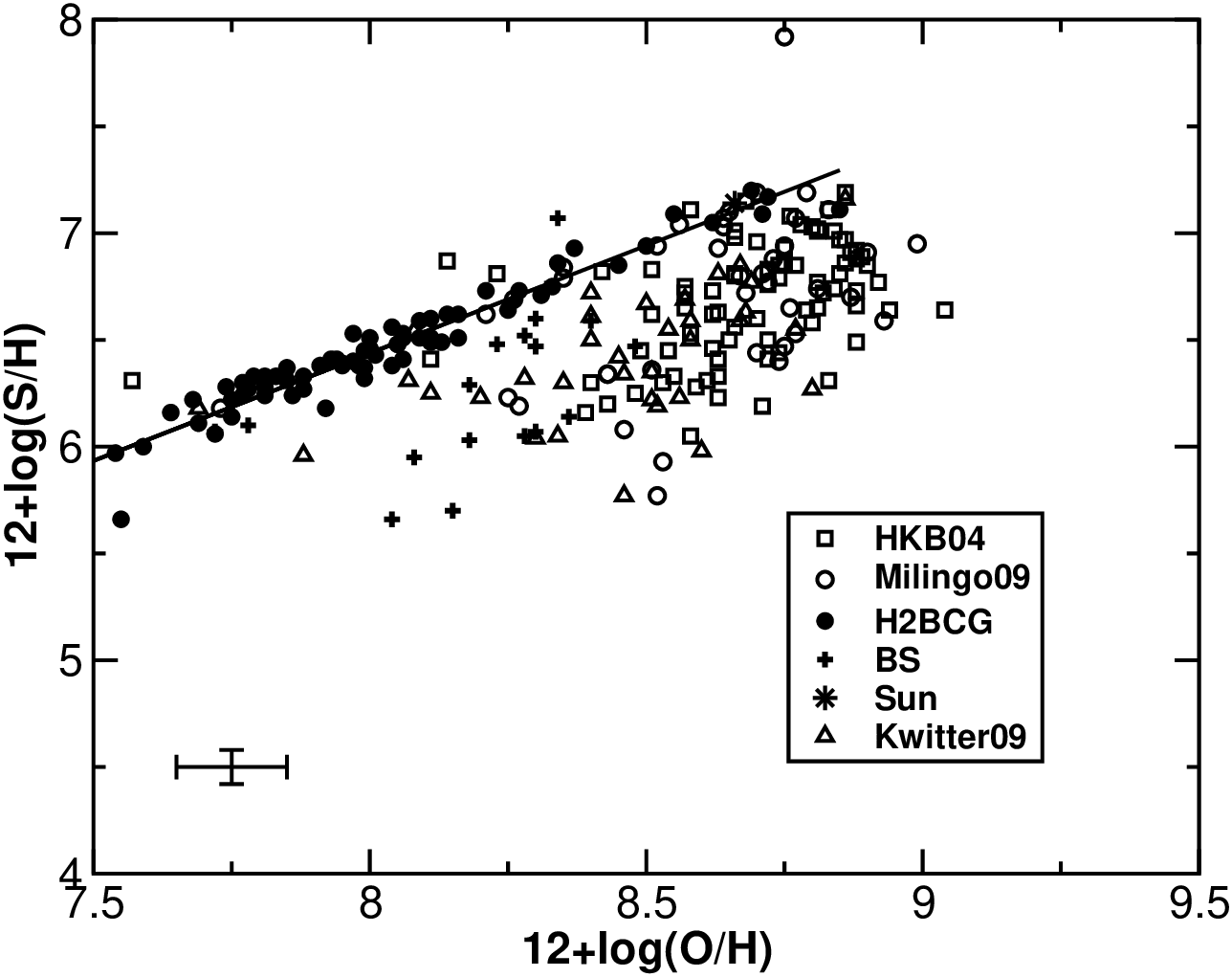} 
   \caption{Same as Fig.~\ref{nevo} but for S versus O.}
   \label{svo}
\end{figure}

\begin{figure}
   \includegraphics[width=4in,angle=0]{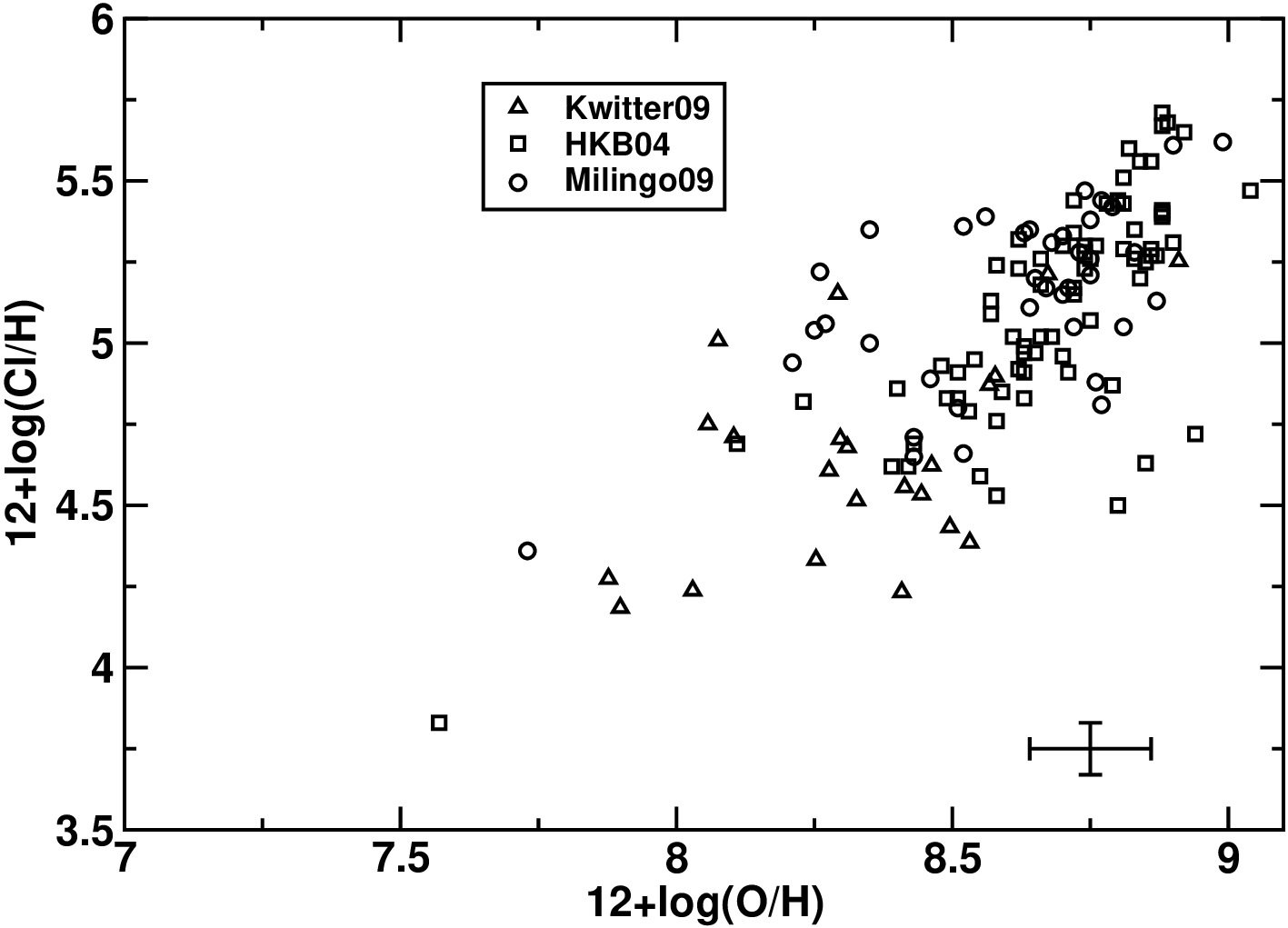} 
   \caption{Same as Fig.~\ref{nevo} but for Cl versus O. No Cl data are available for the H2BCG objects.}
   \label{clvo}
\end{figure}

\begin{figure}
   \includegraphics[width=4in,angle=0]{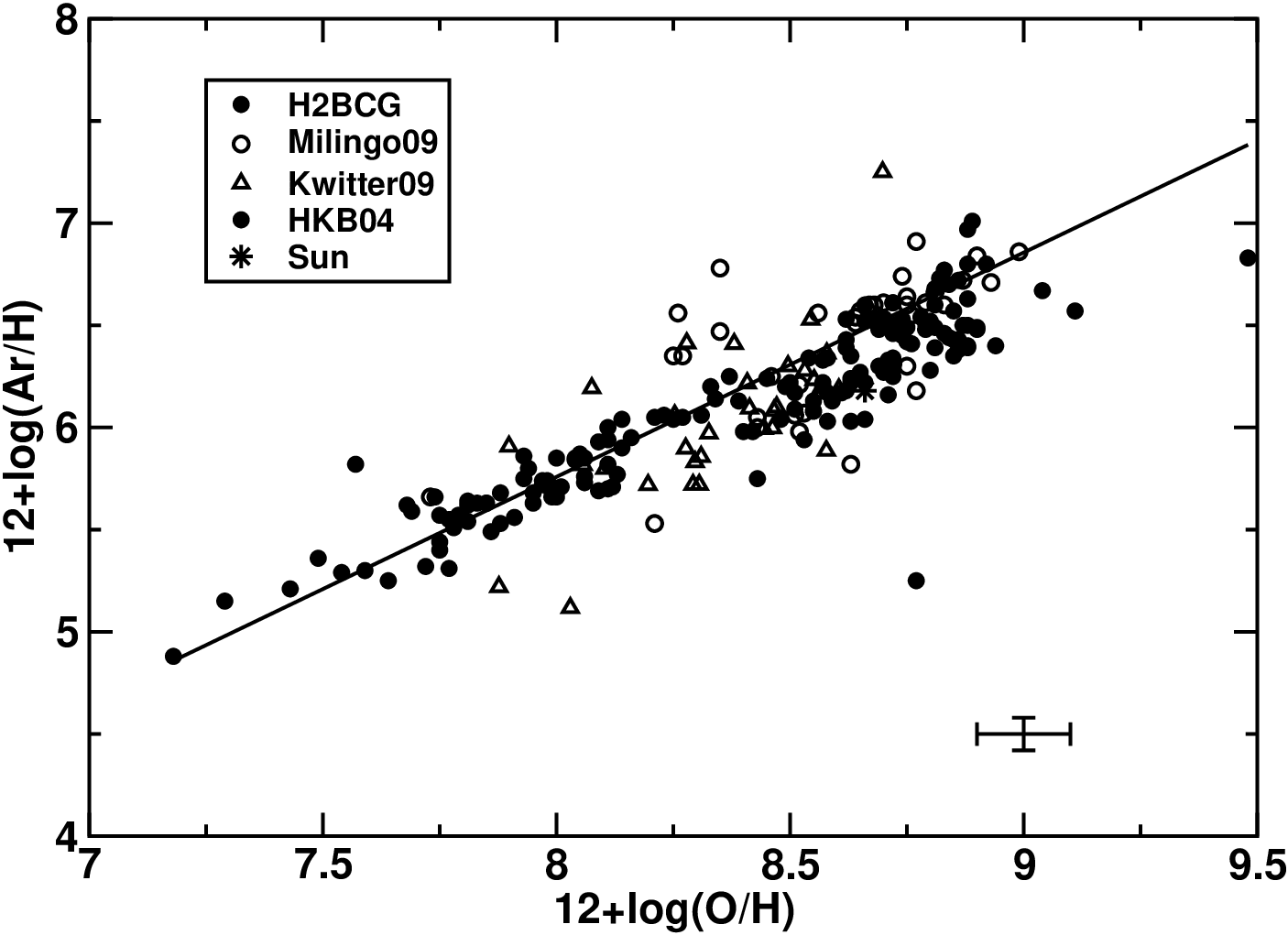} 
   \caption{Same as Fig.~\ref{nevo} but for Ar versus O.}
   \label{arvo}
\end{figure}

\begin{figure}
   \includegraphics[width=4in,angle=0]{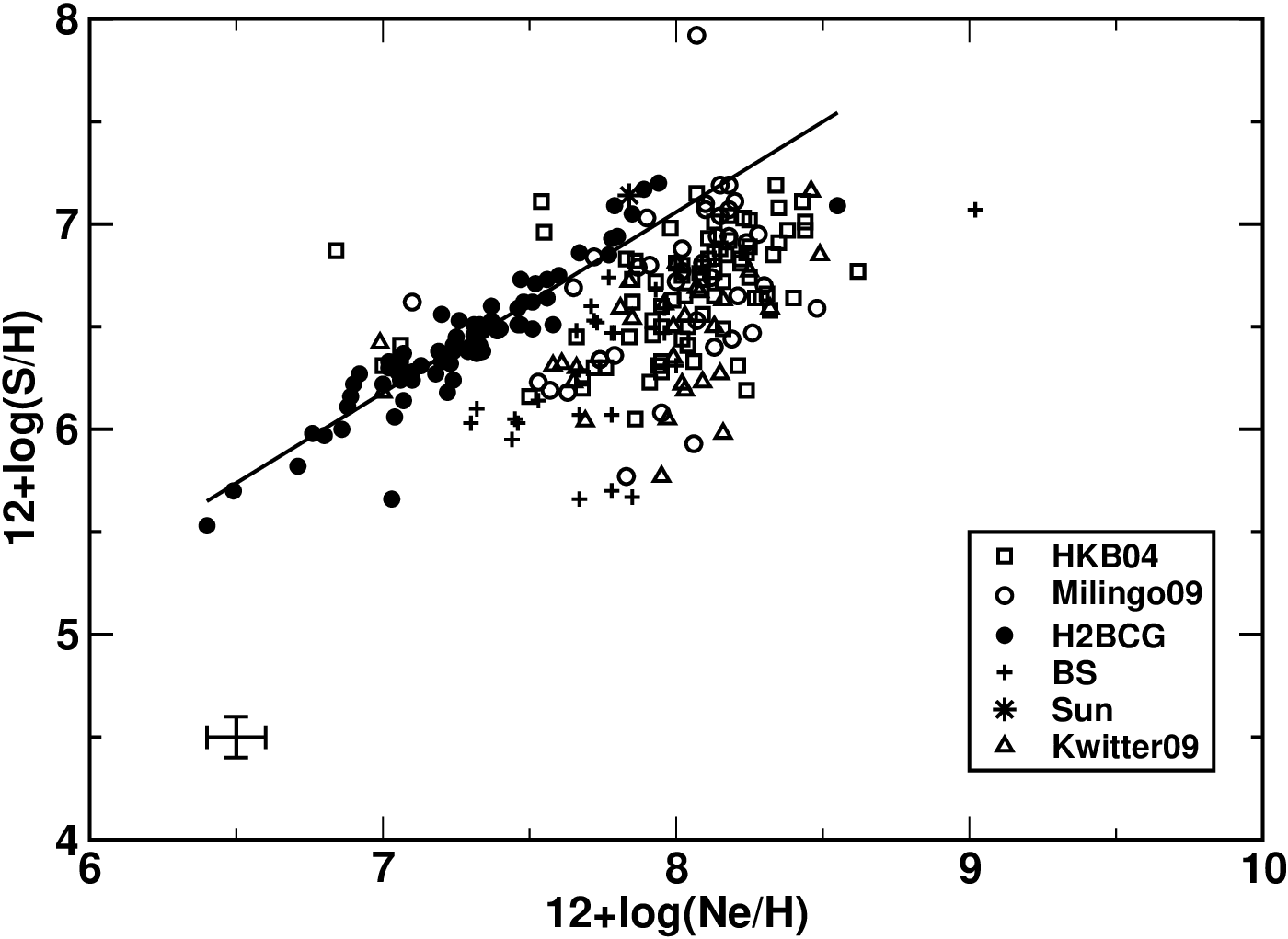} 
   \caption{Same as Fig.~\ref{nevo} but for S versus Ne.}
   \label{svne}
\end{figure}

\begin{figure}
   \includegraphics[width=4in,angle=0]{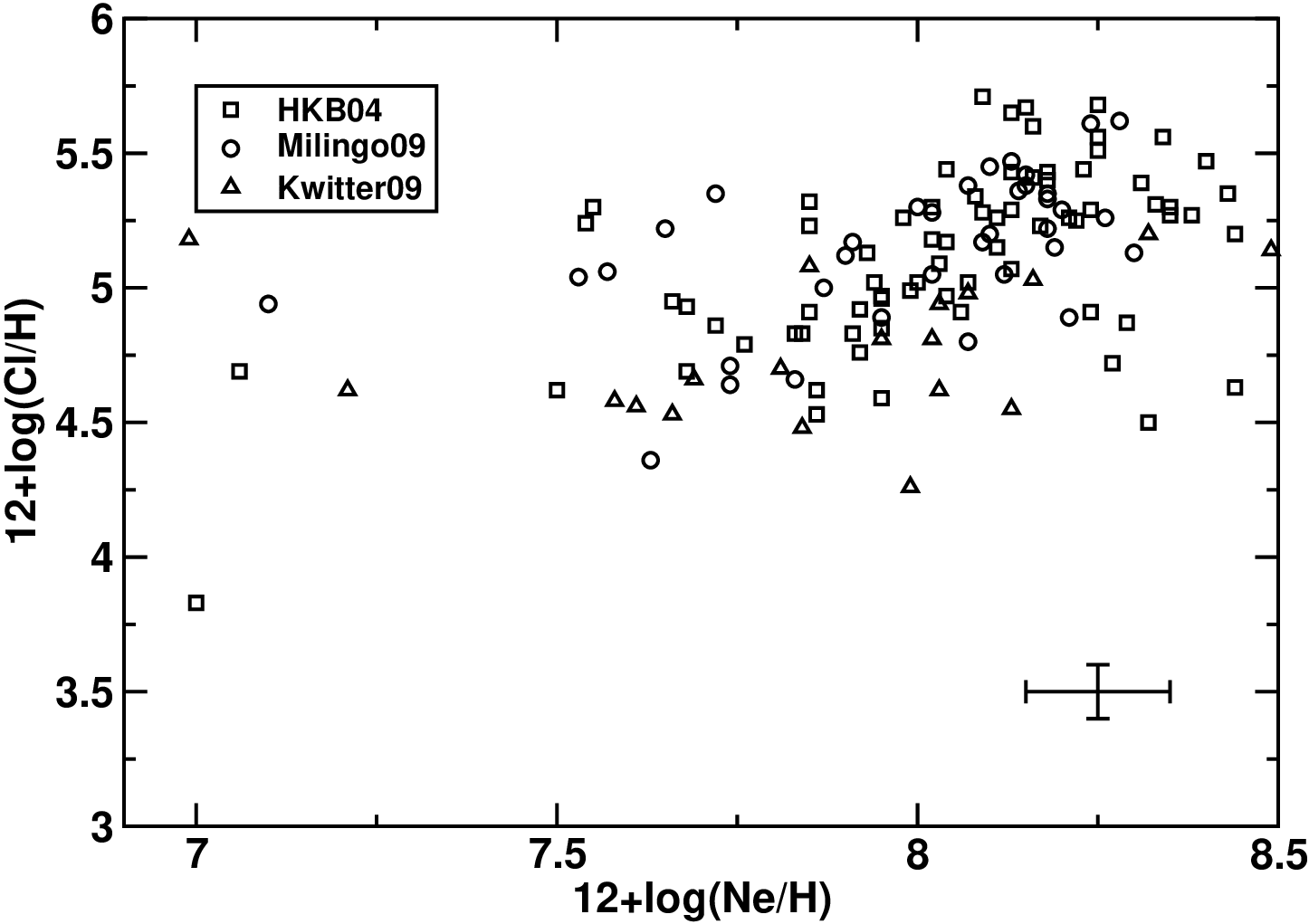} 
   \caption{Same as Fig.~\ref{clvo} but for Cl versus Ne.}
   \label{clvne}
\end{figure}

\begin{figure}
   \includegraphics[width=4in,angle=0]{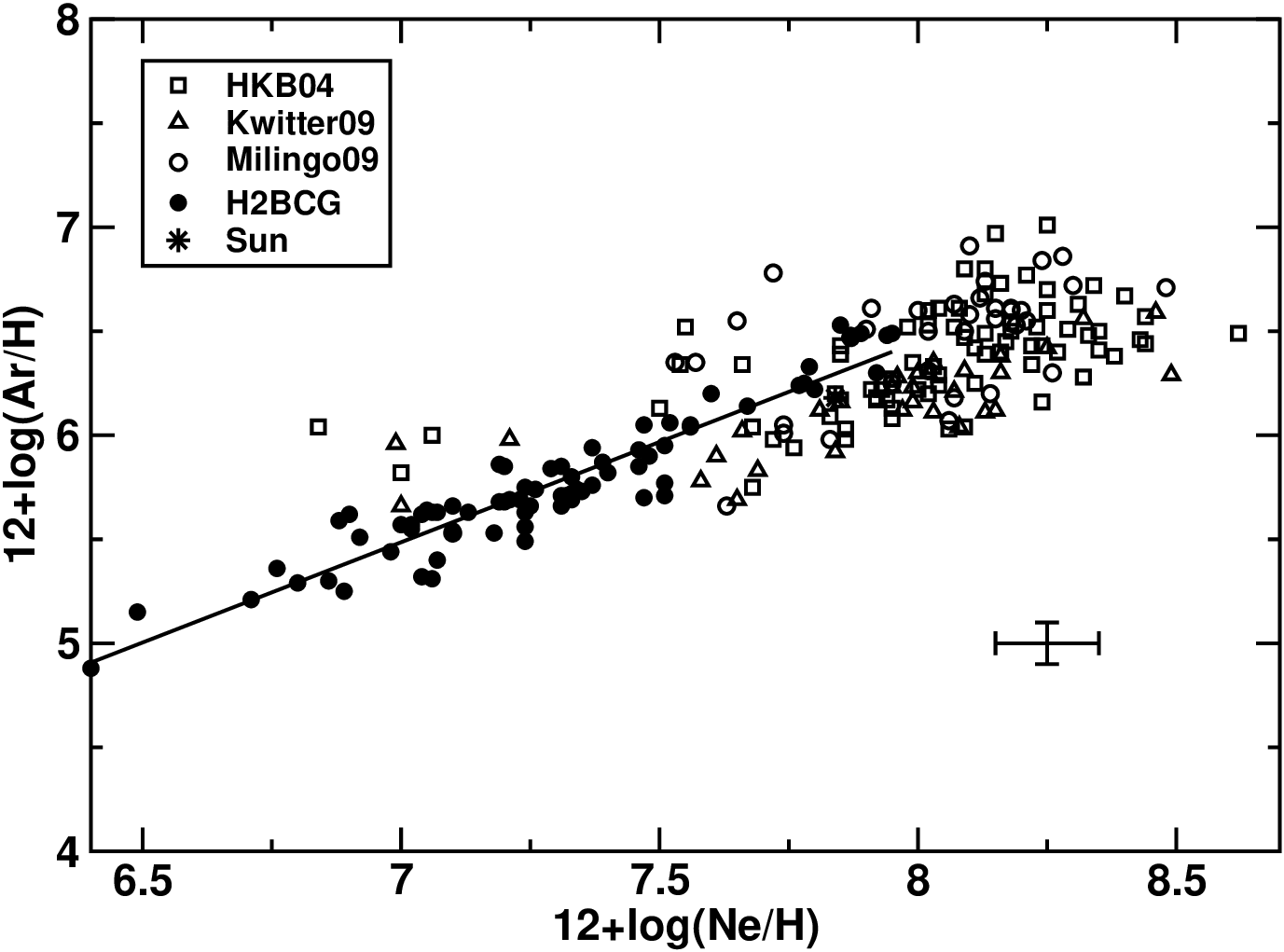} 
   \caption{Same as Fig.~\ref{nevo} but for Ar versus Ne.}
   \label{arvne}
\end{figure}

\begin{figure}
   \includegraphics[width=4in,angle=0]{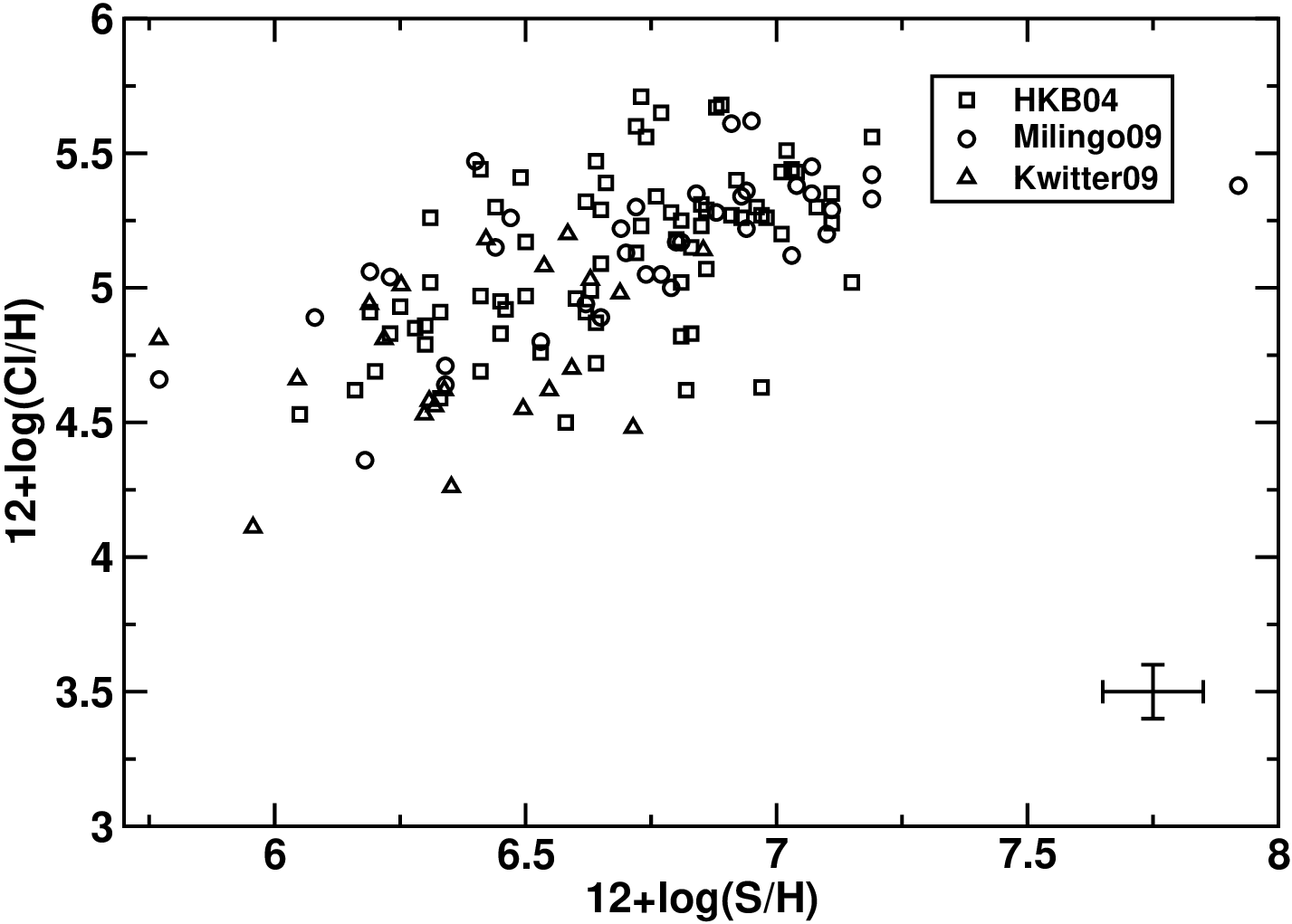} 
   \caption{Same as Fig.~\ref{clvo} but for Cl versus S.}
   \label{clvs}
\end{figure}

\begin{figure}
   \includegraphics[width=4in,angle=0]{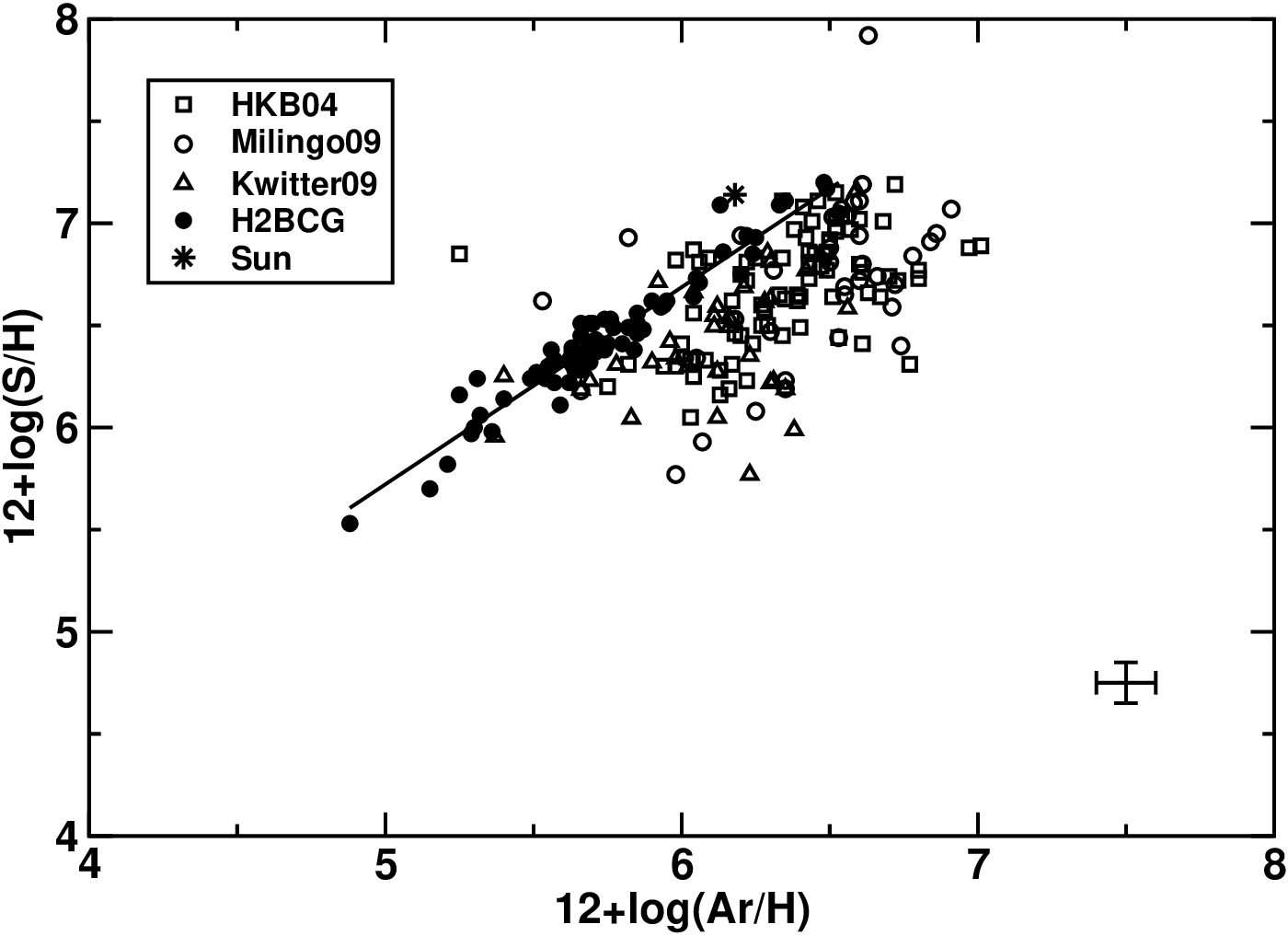} 
   \caption{Same as Fig.~\ref{nevo} but for S versus Ar.}
   \label{svar}
\end{figure}

\begin{figure}
   \includegraphics[width=4in,angle=0]{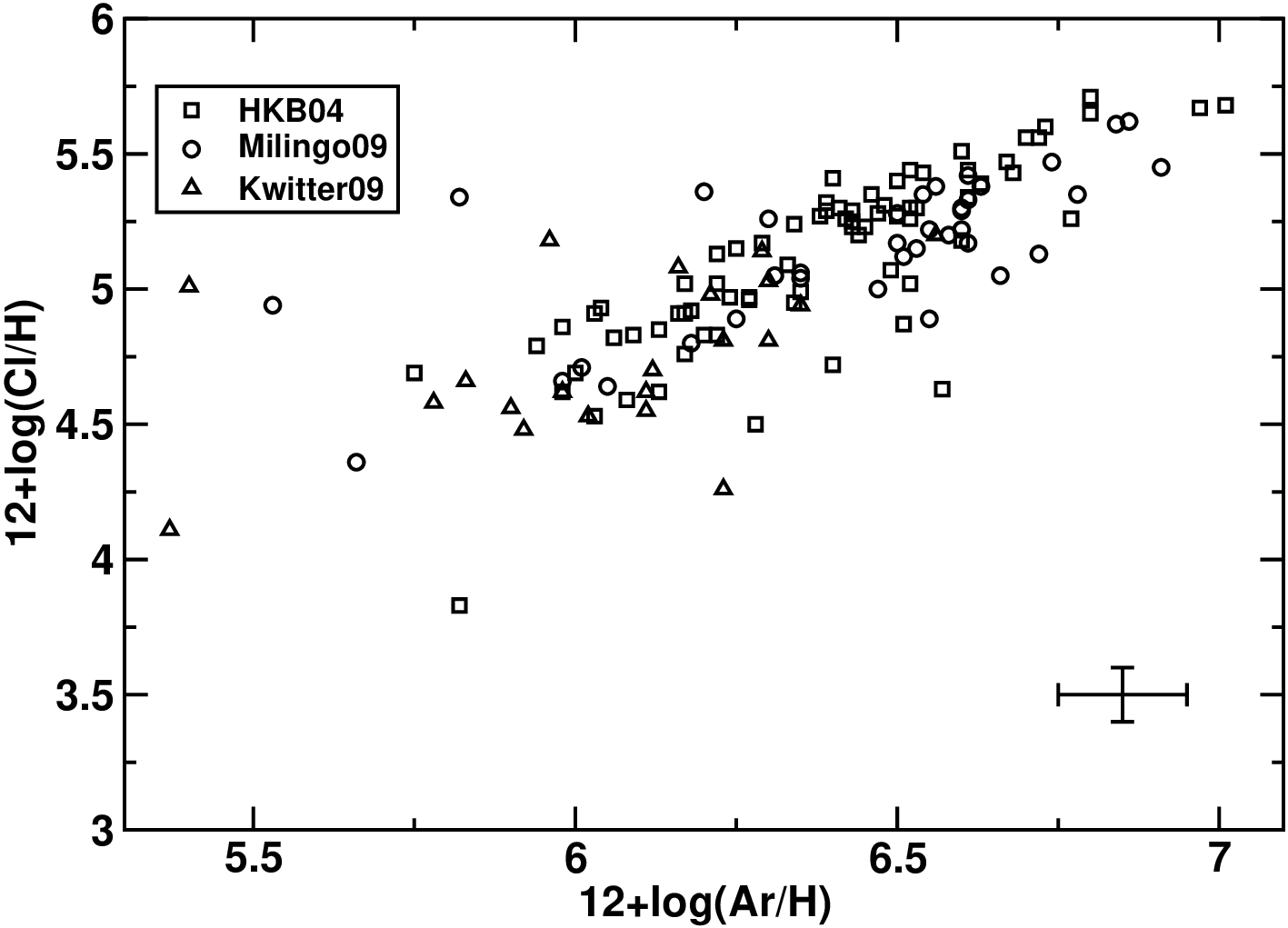} 
   \caption{Same as Fig.~\ref{clvo} but for Cl versus Ar.}
   \label{clvar}
\end{figure}

We now proceed to study PN abundance patterns with respect to the established H2BCG baselines contained in Figs.~\ref{nevo}-\ref{clvar}. In doing so, we note that three of the four plots involving sulfur, Figs.~\ref{svo}, \ref{svne}, and \ref{svar}, display what we have previously described in HKB04 as the sulfur anomaly, i.e., the systematic tendency for PN sulfur abundances to be less than those of H2BCG at the same metallicity. This topic is addressed in a separate subsection. We presently focus on the PN distribution in the remaining six plots and make comparisons with the abundance distributions of the H2BCG objects where possible.

We carried out a linear regression analysis on the PN data in Figs.~\ref{nevo}-\ref{clvar} for the six cases not involving sulfur. The resulting parameters are provided in the bottom portion of Table~\ref{lsf}. While in the first three cases, Ne-O, Cl-O, and Ar-O, the slopes are consistent with unity, the slopes for Cl-Ne, Ar-Ne, and Cl-Ar are not. In fact the slopes of the Cl vs. Ne and Ar vs. Ne relations differ significantly from unity. If we compare the slopes of PNe and H2BCG in Table~6 for the same relations, we see that agreement is present in the cases of Ne-O and Ar-O but not in the case of Ar-Ne. Here the PN slope (+0.57) is much shallower than the slope for H2BCG (+0.96). We suspect that the flatter slope for PNe is due to the six low metallicity PNe residing above the H2BCG track in Fig.~\ref{arvne}. By excluding these objects one sees that the slope would be closer to unity. The slopes of the other two element relations which differ from unity, Cl-Ne and Cl-Ar, are likely influenced by the greater uncertainty in the Cl abundances, given the weak line strengths of the Cl ions in our spectra. Thus, after this initial look PN abundance patterns appear to be reasonably consistent with the expectations of lockstep behavior predicted by theory. If any of the alpha elements are enhanced or depleted by PN progenitor stars, the effect is too small to manifest itself in these plots. We now take a closer look at O, Ne, and Ar.

\begin{figure}
   \includegraphics[width=4in,angle=0]{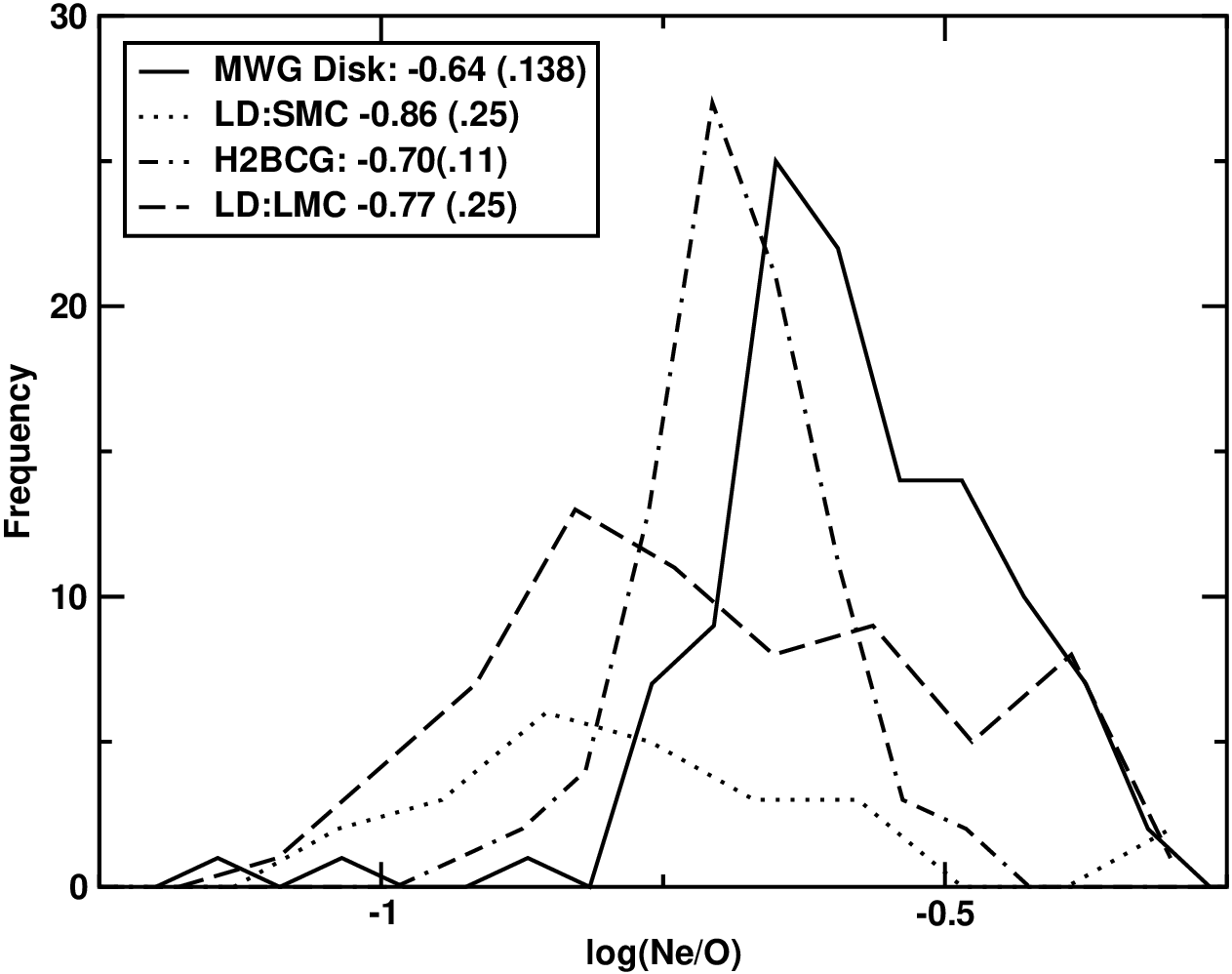} 
   \caption{Frequency distribution of log(Ne/O) for the Galactic PNe in this study, the PNe in the LMC and SMC from \citet{LD06}, and for H2BCGs.  The numbers in the legend represent the average of the logs (and the standard deviation).}
   \label{ne2o_hist}
\end{figure}

\begin{figure}
   \includegraphics[width=4in,angle=0]{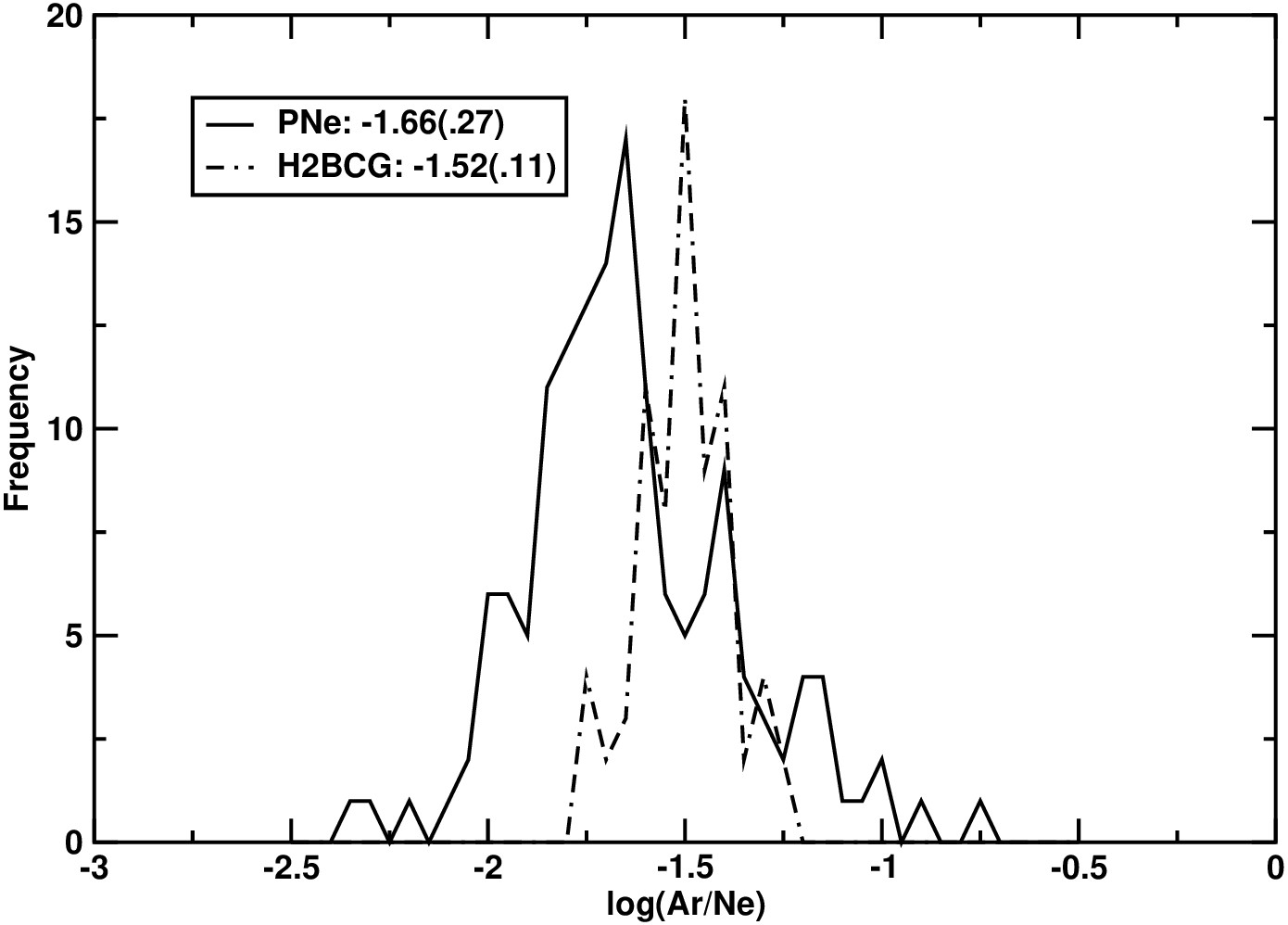} 
   \caption{Same as Fig.~\ref{ne2o_hist} but for log(Ar/Ne)}
   \label{ar2ne_hist}
\end{figure}

\begin{figure}
   \includegraphics[width=4in,angle=0]{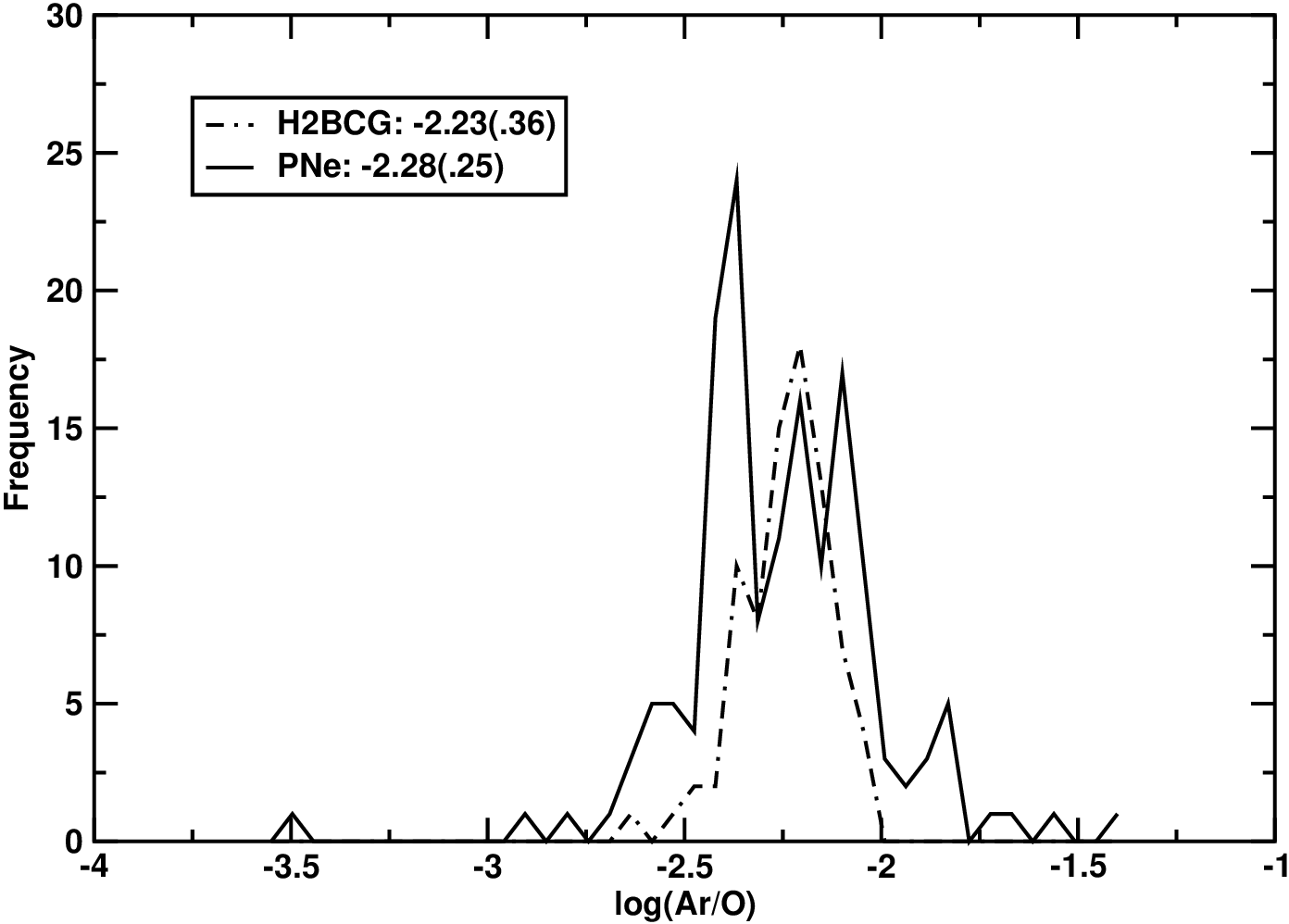} 
   \caption{Same as Fig.~\ref{ne2o_hist} but for log(Ar/O)}
   \label{ar2o_hist}
\end{figure}

{\it O, Ne, and Ar}: We searched for potential differences between PNe and H2BCG by forming ratios of Ne/O, Ar/Ne, and Ar/O, and then comparing the PN and H2BCG frequency distributions in each of the three instances. Such histograms could potentially reveal significant differences between the two object types if nuclear processing is occurring in the PN progenitors. The resulting histograms are displayed in Figs.~\ref{ne2o_hist}-\ref{ar2o_hist}. The numbers in the legend give the average and standard deviation of the log of the relevant element ratio. 

Fig.~\ref{ne2o_hist} shows that Ne/O in PNe is slightly higher on average than in H2BCGs although the difference in the averages is only 0.06~dex, much smaller than the dispersion in the values for PNe. More importantly, while the H2BCG distribution is symmetrical, the PN distribution is not, being noticeably more extended in the direction of higher values of Ne/O and consistent with the tendency in Fig.~\ref{nevo} for PNe to fall above the H2BCG trend line. Also shown in Fig.~\ref{ne2o_hist} are distributions for the LMC and SMC PNe reported in \citet{LD06}. Since the order of increasing metallicity for the four samples is SMC, LMC, H2BCG, MW disk PNe, there is an indication here that Ne/O in PNe increases with metallicity, in agreement with the findings of \citet{WL08}. While temperature-sensitive lines such as [O~III] $\lambda$4363 generally become weaker and more difficult to measure as metallicity rises (due to more efficient cooling) a simple numerical exercise reveals that the observed positive offset in Ne/O from the H2BCG line would require that temperature be in error by several thousand degrees below the real temperature. Since in general weak lines are usually measured to be stronger than they really are, that would push $\lambda$4363 and the corresponding [O~III] temperature up, i.e., in the opposite direction necessary to explain the high values of Ne/O. We can therefore eliminate temperature measuring errors as the source of the systematically high Ne/O derived for PNe.

The average value for log(Ar/Ne) in PNe, according to Fig.~\ref{ar2ne_hist}, is less than that of H2BCGs by 0.14~dex, and again the distribution is broader for PNe, with the extension this time appearing on the low side. In this case the offset may be significant, since it is roughly one standard deviation in magnitude.  Finally, Fig.~\ref{ar2o_hist} indicates that PNe on average have slightly lower values of log(Ar/O), although the difference is very small. At the same time, we note that now the PN distribution appears symmetrical.

Thus, compared with H2BCGs, Ne/O in PNe seems to extend to higher values, Ar/Ne extend to lower values, and the Ar/O distribution appears very similar. Taken together these three histograms suggest that some Ne enhancement is present in a significant portion of PNe, while there is no evidence for similar shifts in Ar and O. If correct, this is an important result, as we can now say that the elevated Ne/O ratio in PNe is due specifically to Ne enhancement and not O depletion. Furthermore, it suggests that LIMS, the stellar progenitors of PNe, actually produce some Ne during their evolution.

\begin{figure}
   \includegraphics[width=4in,angle=0]{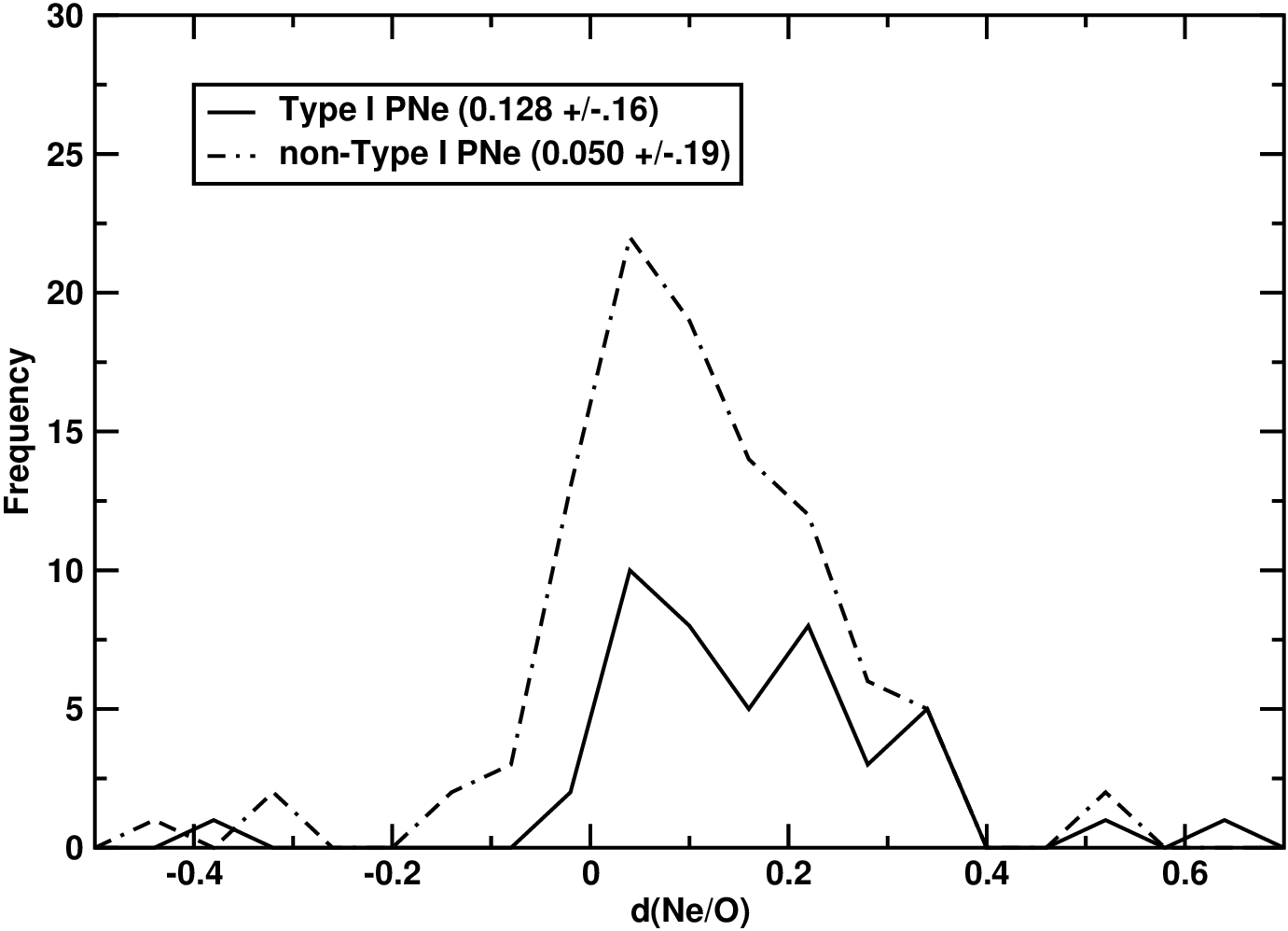} 
   \caption{Frequency distribution of log(Ne/O) differential offsets for Type I PNe and non-Type I PNe in our sample. The numbers in the legend indicate the average of the log values and the standard deviations.}
   \label{dne2o_histo}
\end{figure}

\begin{figure}
  \includegraphics[width=4in,angle=0]{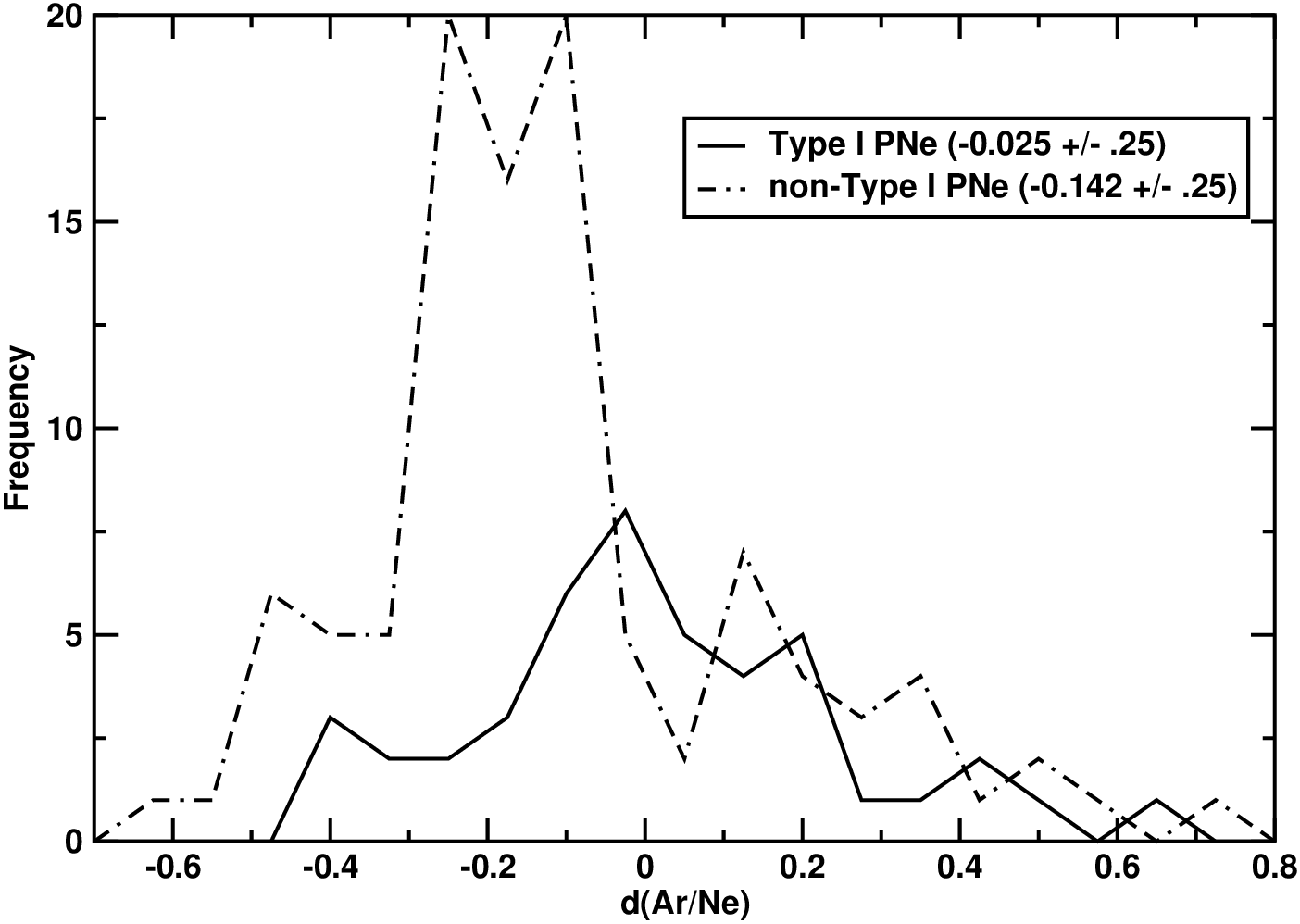} 
  \caption{Same as Fig.~\ref{dne2o_histo} but for Ar/Ne.}
  \label{dar2ne_histo}
\end{figure}

One of the primary motivations for this study was to look for differences between Peimbert Type~I and non-Type~I PNe in terms of alpha element nucleosynthesis in their progenitor stars. To explore this question, we considered the offsets of PNe from H2BCG samples for the two element pairs Ne-O and Ar-Ne. We first computed Ne/O and Ar/Ne for all objects, determined a least squares fit to the H2BCG ratios in both cases (reported in the last two entries of the top panel in Table~\ref{lsf}) and then calculated PN offsets, i.e., d(Ne/O) and d(Ar/Ne).  In each case, for element ratio X, $d(X)=X_{PN}-X_{H2BCG}$. We then separated the PN sample into Type~I and non-Type-I objects to see if there is a difference between these types in terms of the offsets d(Ne/O) and d(Ar/Ne) from H2BCGs. Figs.~\ref{dne2o_histo} and \ref{dar2ne_histo} display these comparisons.  In both figures the two PN types appear to overlap closely and their averages differ by less than a standard deviation. Thus, we see no compelling evidence to suggest that Type~I and Type~II progenitors differ with respect to their distributions of their offsets from the H2BCG sample. This conclusion is supported theoretically by recent work of \citet{KA09}, who find that Ne is only produced in low and intermediate mass stars over a very narrow mass range.  A significant difference between PN types is not expected in regards to alpha element nucleosynthesis.

In conclusion, we have presented evidence to support the idea that many PN progenitors possess enhanced levels of Ne compared to H2BCG, although O and Ar appear to remain unchanged during the star's evolution.  If this Ne enhancement is real it is marking in situ production of $^{22}$Ne and the TDU in action \citep{KL2003, Busso2006}.  However, we see no difference between Type~I and non-Type~I PNe in terms of alpha element nucleosynthesis, despite the fact that the former exhibit enhanced N/O and He/H which are familiar empirical signatures of nuclear processing in LIMS.  This point is emphasized by our finding that Type I and Type II PNe, which differ greatly in N/O, demonstrate no apparent difference in Ne/O.

We note the presence of a small group of outliers especially in the case of Ne and O in Fig.~\ref{nevo}. Their deviation from the norm indicates that in some cases either alpha destruction or production, particularly in the case of Ne, does in fact occur, or alternatively that these objects were formed out of unmixed interstellar materials such that their composition does not accurately sample their current mixed environments. Furthermore, we caution that the presence of those same outliers may help to hide systematic trends such as an increase in Ne with metallicity. After all, it is difficult to decide the guidelines for labeling an object an outlier and subsequently removing it from the sample when we are searching for systematic trends. It is possible that subtle trends connected with alpha element evolution do exist in PNe but are concealed by the uncertainties in our abundance determinations or the presence of outliers.

Finally, we find that Ne/O is systematically higher in the general PN population compared to H2BCGs within the range of 8.5-9.0 for 12+log(O/H). This is consistent with the finding of \citet{WL08}.  

\subsection{The Sulfur Anomaly}

We noted previously that PN sulfur abundances appear to be significantly offset from the H2BCG track in Figs.~\ref{svo}, \ref{svne}, and \ref{svar}. In addition, the scatter in S abundances for a given metallicity is significantly larger than in the case of H2BCG. This situation was first recognized by \citet{HKB04} who dubbed it the Ôsulfur anomalyÕ, i.e., the systematic tendency for PNe to possess lower S abundances than do H~II regions of similar oxygen abundance (metallicity) for reasons which have yet to be confirmed. Given our findings above concerning the similarities between PN and H2BCG abundance patterns for the alpha elements O, Ne, and Ar, we would expect S, another alpha element, to vary in lockstep with the others.

\begin{figure}
   \includegraphics[width=4in,angle=0]{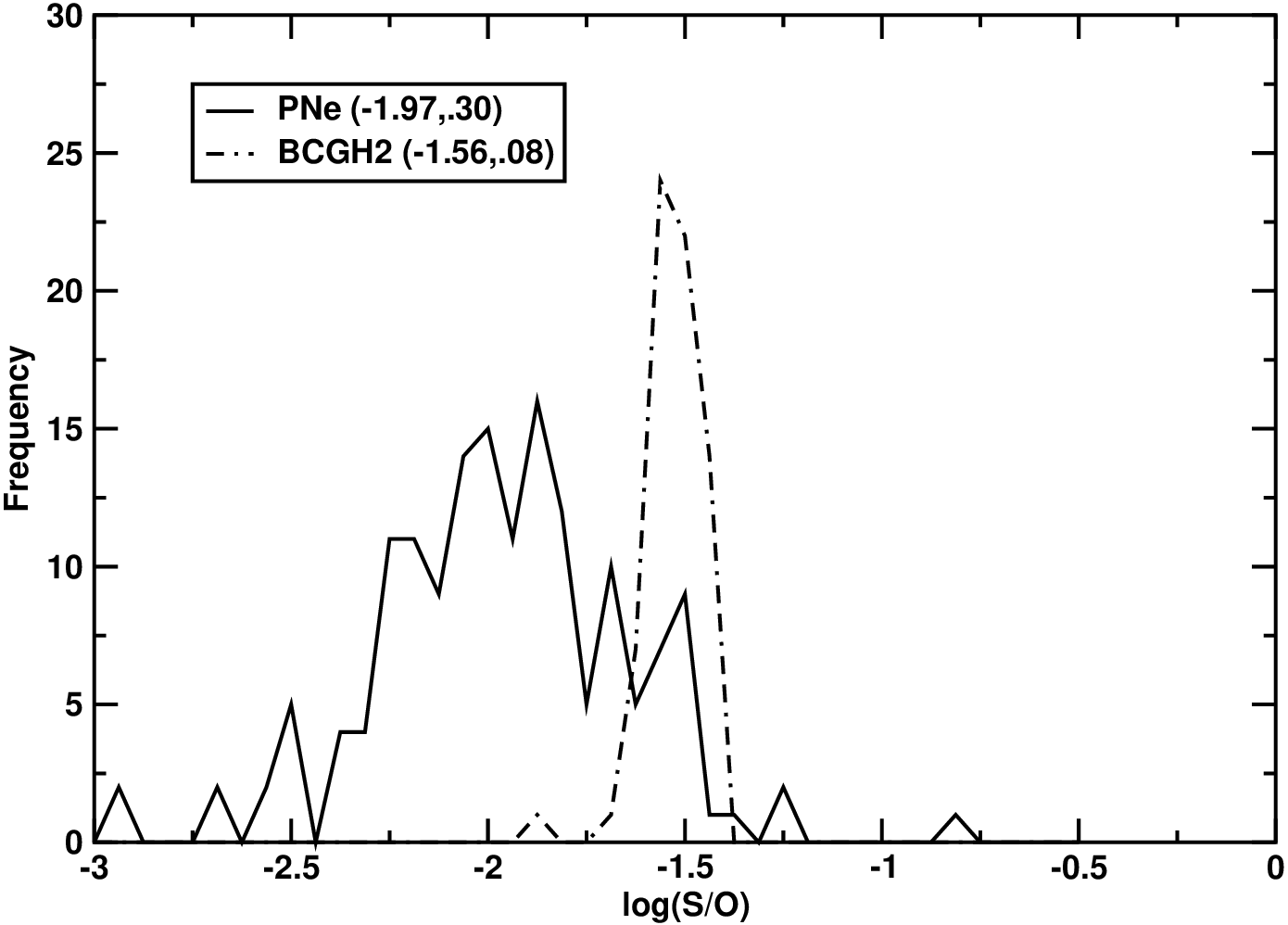} 
   \caption{Frequency distribution of log(S/O) for our PN sample and H2BCG. Averages of the logs and standard deviations appear in the legend.}
   \label{s2o_hist}
\end{figure}

\begin{figure}
   \includegraphics[width=4in,angle=0]{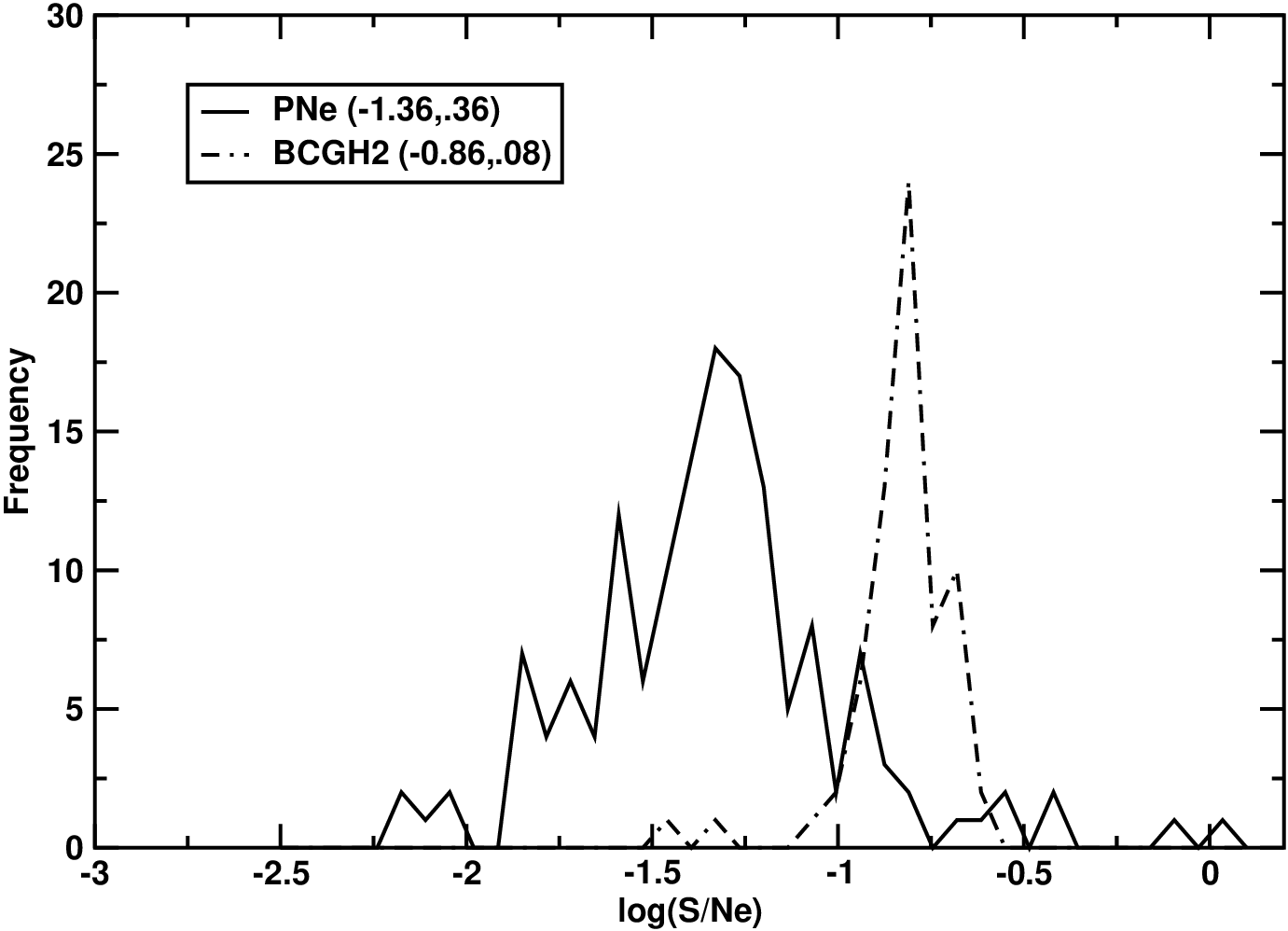} 
   \caption{Same as Fig.~\ref{s2o_hist} but for log(S/Ne).}
   \label{s2ne_hist}
\end{figure}

\begin{figure}
   \includegraphics[width=4in,angle=0]{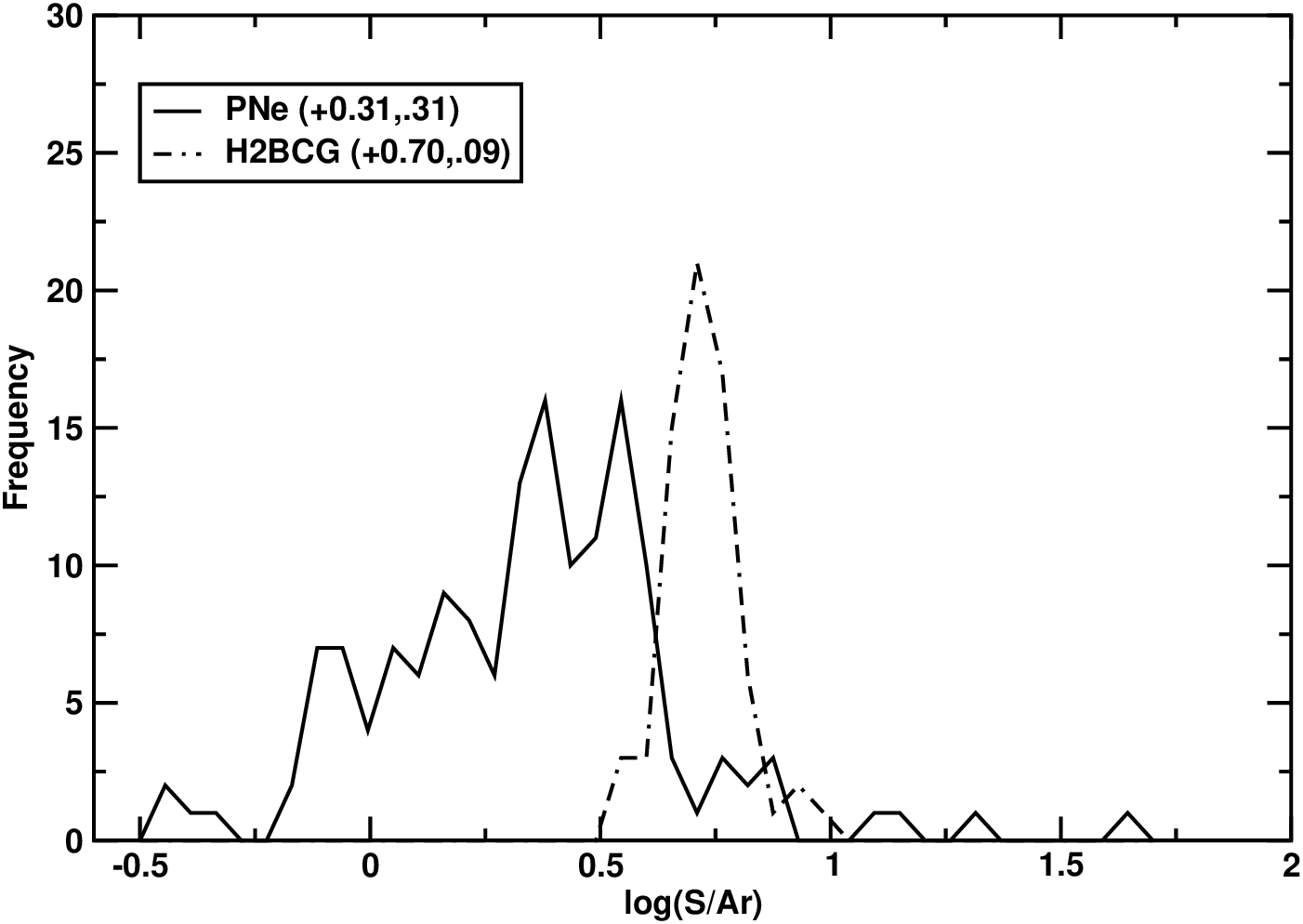} 
   \caption{Same as Fig.~\ref{s2o_hist} but for log(S/Ar).}
   \label{s2ar_hist}
\end{figure}

An appreciation of the sulfur anomaly can be obtained by comparing the frequency distributions of log(S/O), log(S/Ne), and log(S/Ar) for our PN sample and the H2BCG objects (Figs.~\ref{s2o_hist}-\ref{s2ar_hist}).  The average of the logs along with the standard deviation is given in the legend in each case. In contrast to the large overlap visible in the comparisons of the other alpha elements above, there is a clear distinction between the two object groups in each of the three graphs.  Comparing the averages of the logs of the abundance ratios we see that the offsets between the PNe and H2BCG are roughly 0.45~dex, with the PN average always falling substantially below that for H2BCG. In addition, the standard deviations indicate that the spread in each ratio is much greater for PNe than for H2BCG.

In the original discussion of this situation in \citet{HKB04} it was proposed that sulfur anomaly was likely related to problems inherent in the determination of the ICF for sulfur, since photoionization models indicate that in many PNe a significant amount of S is in the S$^{+3}$ ionization stage, which is inaccessible at visual wavelengths. During the intervening time, however, numerous PNe have been studied spectroscopically in the IR by researchers using the Infrared Space Observatory and Spitzer Space Telescope, both of which provide access to the [\ion{S}{4}] emission line centered at 10.5$\mu$m.  These data obviate the need for using a sulfur ICF and provide for the direct computation of a total sulfur abundance simply by adding together the abundances of S$^+$, S$^{+2}$, and S$^{+3}$.

In Figs.~\ref{svo} and \ref{svne} we have included the PN sulfur abundances (triangles) from \citet{BS08} for a small sample of PNe whose S abundances were determined using IR data in the manner just described. One can readily see that the sulfur anomaly persists, as the scatter and systematically low S abundances are similar to what we find in our own sample. Therefore,  we can now state with some confidence that the sulfur anomaly is apparently unrelated to the sulfur ICF.

It has been suggested by \citet{PBS06} that the sulfur anomaly could be the result of dust formation. In particular, S may be removed by the formation of compounds such as MgS and FeS in those PNe exhibiting large S deficits. This would be expected to occur more readily in C-rich environments, where sulfide formation is favored \citep{Lattimer78, Zhang09}. Initial plots of the deviation of the S abundance from the expected value determined by the H2BCG track versus C/O shows essentially no trend, where one might expect a correlation if a larger C/O is associated with C-rich environments where MgS can form. In addition, a simple calculation shows that the average S deviation would correspond to a drop of about 0.1~dex if in fact the sulfur deficit were due to MgS dust formation, an offset which is likely undetectable. 

\section{Summary/Conclusions}
Continuing our previous work we present line strengths, abundances, and X/O element ratios for a sample of 38 Galactic disk PNe bringing the total to $>$ 120.  The extended spectral coverage for this large sample of objects is atypical making available the [\ion{S}{3}] lines at 9069$\AA$ and 9532$\AA$ for use in determining nebular diagnostics as well as total sulfur abundance.  Including these objects in our compilation we've attempted to discern signatures of alpha element processing in PNe progenitor stars. We conclude the following:

\begin{enumerate}

\item A careful comparison of the behavior of Ne, O, and Ar with respect to each other in PNe and H2BCG reveals systematically higher values of Ne/O and lower values of Ar/Ne in PNe within the range of 8.5-9.0 of 12+log(O/H), suggesting that Ne is enhanced in many of these objects. The enhancement is likely coming from the extended processing of $^{14}$N by alpha capture to yield $^{22}$Ne.

\item The positions of numerous outliers in several plots cannot be explained away by uncertainties in the abundance determinations. The progenitors of these objects may have formed from poorly mixed interstellar material or show significant synthesis or depletion of alpha elements.

\item Type I and non-Type~I PNe exhibit no difference in their alpha element abundance patterns.

\item The sulfur anomaly, in which sulfur is found to be systematically lower in PNe than in H2BCG at the same metallcity, is reconfirmed. Recent work shows that the sulfur anomaly is likely not related to problems with the ICF and at first glance is too large to be accounted for by dust formation.
\end{enumerate}

\acknowledgments
The authors gratefully acknowledge support from the Bronfman Science Center at Williams College, NSF grants \#AST-0806577,  \#AST-0806490 $\&$ \#AST-9819123 to the University of Oklahoma and Williams College, the Faculty Development Committee at Gettysburg College, and the Pennsylvania Space Grant Consortium.   We thank the numerous students involved in this project including Kate Barnes (Franklin $\&$ Marshall '06), Johanna Teske (American University '08), Davis Stevenson (Williams '04), Joseph Gangestad (Williams '06), Anne Jaskot (Williams '08), and Henry Bradsher (Oklahoma '08).  We thank Reggie Dufour, Bob Rubin, and Fabio Bresolin for their input concerning neon abundances in metal-rich H~II regions, John Cowan for conversations about Cl production, and finally Bruce Balick and the referee for carefully reading over and commenting on the manuscript.

\pagebreak
\appendix

\begin{deluxetable}{lcccc}
\tablecolumns{5} \tablewidth{0pc} \tabletypesize{\scriptsize}
\tablenum{1} \tablecaption{Observation Information} \tablehead{
\colhead{Object} & \colhead{Offset
($\arcsec$)\tablenotemark{1}} & \colhead{Total Blue Exp (sec)}&
\colhead{Total Red Exp (sec)} & \colhead{Observatory\tablenotemark{2}}}
 
\startdata
Cn3-1 &\nodata &429 &480 &K04 \\
H2-18 &\nodata &1200 &1800 &K04 \\
Hb4 &\nodata &1800 & 1600 &K03 \\
Hb6 &\nodata &800 &600 &K03 \\

He2-111 &\nodata &4260 &2940 &C04 \\
Hu 1-2 &\nodata &1420 &1030 &K03, K04 \\

M1-8 &\nodata &4200 &3600 &C03 \\
M1-13 &\nodata &1620 &2300 &C03 \\
M1-17 &\nodata &960 &1200 &C03 \\
M1-40 &\nodata &1500 &1200 &K03 \\
M1-42 &\nodata &1650 &3360 &C04 \\

M2-52 &\nodata &3600 &5000 &K03 \\
M2-55 &\nodata &1800 &450 &K04 \\
M3-2 &\nodata &3300 &3000 &C03 \\
M3-3 &\nodata &3000 &3600 &C03 \\
M3-5 &\nodata &2100 &2400 &C03 \\
Me1-1 &\nodata &250 &1125 &K03 \\
Me2-2 &\nodata &1400 &1080 &K03, K04 \\
Mz 2 &\nodata &3180 &3612 &C04 \\

Mz 3 &\nodata &1800 &2880 &C04 \\
NGC 1535 &\nodata &2360 &2500 &C03 \\
NGC 2452 &\nodata &1500 &1200 &C03 \\
NGC 5315 &\nodata &961 &855 &C04 \\
 
NGC 6302 &\nodata &700 &1350 &C04 \\
NGC 6369 &10N &2000 &2200 &K03 \\
NGC 6445 &25N &1200 &1200 &K03 \\
NGC 6537 &\nodata &725 &300 &K03 \\
 
NGC 6741 &\nodata &800 &400 &K03 \\
NGC 6751 &6S &1500 &1500 &K03 \\
NGC 6778 &\nodata &1500 &1800 &K03 \\
NGC 6803 &\nodata &300 &150 &K03 \\
NGC 6804 &10S &1800 &5100 &K03 \\
NGC 6853 &83S, 120W &2700 &1800 &K04 \\
NGC 6881 &\nodata &1700 &500 &K03 \\
NGC 7008 &29N, 11E &1200 &1926 &K04 \\

NGC 7354 &\nodata &3503 &4500 &K03 \\
PB6 &\nodata &2000 &3000 &C03 \\

Vy 1-1 &\nodata &360 &1440 &K04 \\
\enddata

\tablenotetext{1}{From central star or center of nebula.}
\tablenotetext{2}{K = KPNO, C = CTIO, number denotes year of run.}
\end{deluxetable}
\clearpage

\begin{deluxetable}{lcr@{}lr@{}lr@{}lr@{}lr@{}lr@{}lr@{}lr@{}lr@{}lr@{}l}
\tabletypesize{\scriptsize}
\setlength{\tabcolsep}{0.07in}
\rotate
\tablecolumns{22}
\tablewidth{0in}
\tablenum{2}
\tablecaption{Fluxes and Intensities}
\tablehead{
\colhead{} & \colhead{} &
\multicolumn{4}{c}{Cn3-1} &
\multicolumn{4}{c}{H2-18} &
\multicolumn{4}{c}{Hb4} &
\multicolumn{4}{c}{Hb6} &
\multicolumn{4}{c}{He2-111} \\
\cline{3-6} \cline{7-10} \cline{11-14} \cline{15-18} \cline{19-22}  \\
\colhead{Line} &
\colhead{f($\lambda$)} &
\multicolumn{2}{c}{F($\lambda$)} &
\multicolumn{2}{c}{I($\lambda$)} &
\multicolumn{2}{c}{F($\lambda$)} &
\multicolumn{2}{c}{I($\lambda$)} &
\multicolumn{2}{c}{F($\lambda$)} &
\multicolumn{2}{c}{I($\lambda$)} &
\multicolumn{2}{c}{F($\lambda$)} &
\multicolumn{2}{c}{I($\lambda$)} &
\multicolumn{2}{c}{F($\lambda$)} &
\multicolumn{2}{c}{I($\lambda$)}}
\startdata
\[[O II] $\lambda$3727 & 0.292 & 145 & & 165 & $\pm$37 & 10.5 & & 24.9 & $\pm$5.55 & 13.6 & & 37.9 & $\pm$8.47 & 13.5 & & 48.7 & $\pm$10.93 & 223 & & 416 & $\pm$95 \\
He II + H11 $\lambda$3770 & 0.280 & 2.73 & & 3.10 & $\pm$0.68 & \nodata & & \nodata&  & \nodata & & \nodata&  & \nodata & & \nodata&  & \nodata & & \nodata&  \\
He II + H10 $\lambda$3797 & 0.272 & 3.36 & & 3.80 & $\pm$0.82 & \nodata & & \nodata&  & \nodata & & \nodata&  & \nodata & & \nodata&  & 2.87 &: & 5.15 & $\pm$1.85: \\
He II + H9 $\lambda$3835 & 0.262 & 5.15 & & 5.79 & $\pm$1.24 & \nodata & & \nodata&  & 2.61 & & 6.54 & $\pm$1.40 & 1.70 &:: & 5.36 & $\pm$2.87:: & 4.25 & & 7.44 & $\pm$1.62 \\
\[[Ne III] $\lambda$3869 & 0.252 & \nodata & & \nodata&  & 41.2 & & 86.8 & $\pm$18.24 & 38.4 & & 93.0 & $\pm$19.61 & 31.5 & & 95.5 & $\pm$20.17 & 150 & & 258 & $\pm$55 \\
He I + H8 $\lambda$3889 & 0.247 & 10.8 & & 12.0 & $\pm$2.50 & 4.36 & & 9.04 & $\pm$1.89 & 5.40 & & 12.8 & $\pm$2.68 & 3.90 &:: & 11.5 & $\pm$6.14:: & 16.2 & & 27.4 & $\pm$5.83 \\
\[[Ne III] $\lambda$3968 & 0.225 & \nodata & & \nodata&  & 10.4 &\tablenotemark{a} & 20.1 & $\pm$7.26\tablenotemark{a} & 12.2 &\tablenotemark{a} & 26.8 & $\pm$8.62\tablenotemark{a} & 9.96 &\tablenotemark{a} & 26.7 & $\pm$8.65\tablenotemark{a} & 42.5 &\tablenotemark{a} & 68.8 & $\pm$17.54\tablenotemark{a} \\
H$\epsilon$ $\lambda$3970 & 0.224 & 13.7 & & 15.1 & & 8.21 &\tablenotemark{a} & 15.9 &\tablenotemark{a} & 7.27 &\tablenotemark{a} & 16.0 &\tablenotemark{a} & 6.03 &\tablenotemark{a} & 16.2 &\tablenotemark{a} & 10.4 &\tablenotemark{a} & 16.9 &\tablenotemark{a} \\
He I + He II $\lambda$4026 & 0.209 & 0.557 &: & 0.612 & $\pm$0.211: & \nodata & & \nodata&  & 0.887 &: & 1.85 & $\pm$0.64: & \nodata & & \nodata&  & 2.04 &:: & 3.19 & $\pm$1.69:: \\
\[[S II] $\lambda$4071 & 0.196 & 3.36 &: & 3.66 & $\pm$1.25: & \nodata & & \nodata&  & 1.71 &: & 3.41 & $\pm$1.17: & 2.09 &:: & 4.95 & $\pm$2.61:: & 12.9 & & 19.6 & $\pm$3.87 \\
\\
\vdots \\
\\
\[[S III] $\lambda$9069 & -0.670 & \nodata & & \nodata&  & 99.7 & & 13.8 & $\pm$3.17 & 510 & & 48.5 & $\pm$11.13 & 471 & & 24.8 & $\pm$5.70 & 256 & & 61.2 & $\pm$13.99 \\
P9 $\lambda$9228 & -0.610 & \nodata & & \nodata&  & 34.6 & & 5.71 & $\pm$1.20 & 37.2 & & 4.37 & $\pm$0.92 & 49.2 & & 3.37 & $\pm$0.71 & 6.74 &: & 1.83 & $\pm$0.64: \\
\[[S III] $\lambda$9532 & -0.632 & 104 & & 78.4 & $\pm$17.08 & 184 & & 28.5 & $\pm$6.19 & 613 & & 66.5 & $\pm$14.45 & 1941 & & 121 & $\pm$26 & 849 & & 220 & $\pm$48 \\
P8 $\lambda$9546 & -0.633 & 8.03 & & 6.05 & $\pm$1.32 & 12.0 & & 1.85 & $\pm$0.40 & 13.1 & & 1.41 & $\pm$0.31 & 61.0 & & 3.78 & $\pm$0.82 & \nodata & & \nodata&  \\
c & & & 0.19 & & &  &1.28 & & & &1.53 &&& &1.91 &&& &0.93 \\
H$\alpha$/H$\beta$ & & & 2.88 & & &  &2.86 &&& &2.87 & &&&2.83 & &&&2.78 \\ 
log F$_{H\beta}$\tablenotemark{b} & & -10.90 & & & & -12.96 & & & & -12.10 & & & & -12.22 & & & & -12.88 \\
\enddata
\tablenotetext{a}{Deblended.}
\tablenotetext{b}{ergs\ cm$^{-2}$ s$^{-1}$ in our extracted spectra}
\end{deluxetable}

\clearpage

\begin{deluxetable}{lr@{}llr@{}llr@{}llr@{}llr@{}ll}
\tabletypesize{\small}
\setlength{\tabcolsep}{0.07in}
\rotate
\tablecolumns{16}
\tablewidth{0in}
\tablenum{3}
\tablecaption{Temperatures and Densities}
\tablehead{
\colhead{} &
\multicolumn{3}{c}{Cn3-1} &
\multicolumn{3}{c}{H2-18} &
\multicolumn{3}{c}{Hb4} &
\multicolumn{3}{c}{Hb6} &
\multicolumn{3}{c}{He2-111} \\
\cline{2-4} \cline{5-7} \cline{8-10} \cline{11-13} \cline{14-16}  \\
\colhead{Parameter\tablenotemark{a}} &
\multicolumn{2}{c}{Value\tablenotemark{b}} &
\colhead{Notes\tablenotemark{c}} &
\multicolumn{2}{c}{Value} &
\colhead{Notes} &
\multicolumn{2}{c}{Value} &
\colhead{Notes} &
\multicolumn{2}{c}{Value} &
\colhead{Notes} &
\multicolumn{2}{c}{Value} &
\colhead{Notes}}
\startdata
T$_{[O~III]}$ & 8490 & $\pm$1134 &  & 9601 & $\pm$438 &  & 9241 & $\pm$408 &  & 11180 & $\pm$1235 &  & 15670 & $\pm$1162 & \\
T$_{[N~II]}$ & 7828 & $\pm$647 &  & 10300 & & Default. & 11720 & $\pm$1274 &  & 12280 & $\pm$1167 &  & 11770 & $\pm$876 & \\
T$_{[O~II]}$ & 7715 & $\pm$2605 &  & 11610 & $\pm$5569 &  & 15120 & $\pm$11090 &  & 18160 & $\pm$14000 &  & 9149 & $\pm$1969 & \\
T$_{[S~II]}$ & 7200 & $\pm$3504 &  & \nodata &  &  & 7612 & $\pm$3824 &  & 11110 & $\pm$8984 &  & 8828 & $\pm$2015 & \\
T$_{[S~III]}$ & 7206 & $\pm$435 & Used 9532.  & 10840 & $\pm$1808 & Used 9069.  & 9988 & $\pm$874 & Used 9069.  & 11380 & $\pm$1075 & Used 9532.  & 15610 & $\pm$1978 & Used 9532. \\
Ne$_{[S~II]}$ & 6195 & $\pm$5072 &  & 4351 & $\pm$2840 &  & 6001 & $\pm$4717 &  & 4552 & $\pm$2952 &  & 583 & $\pm$356 & \\
Ne$_{[Cl~III]}$ & 8451 & $\pm$5980 &  & \nodata &  &  & 5465 & $\pm$5439 &  & 5367 & $\pm$1808 &  & 1238 & $\pm$2726 & \\
\enddata
\tablenotetext{a}{The emission lines (wavelengths in $\AA$) used to calculate electron temperatures and densities were: 5007, 4363 for ${T_{[O~III]}}$; 6584, 5755  for  ${T_{[N~II]}}$; 3727, 7324 for ${T_{[O~II]}}$; 6716, 6731, 4071 for ${T_{[S~II]}}$, 9532, 9069, 6312 for ${T_{[S~III]}}$; 6716, 6731 for ${N_{e, [S~II]}}$; and 5518, 5538 for ${N_{e, [Cl~III]}}$.}  
\tablenotetext{b}{Temperatures and densities given in kelvin and cm$^{-3}$.}
\tablenotetext{c}{If either 5755 or 6584 is unavailable ${T_{[N~II]}}$ is estimated from ${T_{[O~III]}}$ \citep{KH2001}.  Given telluric effects when calculating ${T_{[S~III]}}$ if the ratio of 9532/9069 is $\geq$ 2.48 then 9532 is used as the nebular transition, if the 9532/9069 ratio is less than 2.48 then 9069 is used.  Default ${T_{[N~II]}}$ and ${N_{e, [Cl~III]}}$ values and the high density limit for ${N_{e, [S~II]}}$ are based on criteria discussed in \citet{KH2001}.} 

\end{deluxetable}

\begin{deluxetable}{lcr@{}lcr@{}lcr@{}lcr@{}lcr@{}l}
\tabletypesize{\scriptsize}
\setlength{\tabcolsep}{0.07in}
\rotate
\tablecolumns{16}
\tablewidth{0in}
\tablenum{4}
\tablecaption{Ionic Abundances}
\tablehead{
\colhead{} &
\multicolumn{3}{c}{Cn3-1} &
\multicolumn{3}{c}{H2-18} &
\multicolumn{3}{c}{Hb4} &
\multicolumn{3}{c}{Hb6} &
\multicolumn{3}{c}{He2-111} \\
\cline{2-4} \cline{5-7} \cline{8-10} \cline{11-13} \cline{14-16}  \\
\colhead{Ion} &
\colhead{T$_{\mathrm{used}}$\tablenotemark{a}} &
\multicolumn{2}{c}{Abundance\tablenotemark{b}} &
\colhead{T$_{\mathrm{used}}$} &
\multicolumn{2}{c}{Abundance} &
\colhead{T$_{\mathrm{used}}$} &
\multicolumn{2}{c}{Abundance} &
\colhead{T$_{\mathrm{used}}$} &
\multicolumn{2}{c}{Abundance} &
\colhead{T$_{\mathrm{used}}$} &
\multicolumn{2}{c}{Abundance} 
}
\startdata
He$^{+}$ & [O III] & 5.57 & $\pm$0.70(-2) & [O III] & 0.116 & $\pm$0.015 & [O III] & 0.121 & $\pm$0.015 & [O III] & 0.117 & $\pm$0.016 & [O III] & 0.144 & $\pm$0.021\\
He$^{+2}$ & [O III] & \nodata &  & [O III] & 3.82 & $\pm$1.23(-3) & [O III] & 1.29 & $\pm$0.20(-2) & [O III] & 1.97 & $\pm$0.30(-2) & [O III] & 7.65 & $\pm$1.17(-2)\\
icf(He) &  & 1.00 & &  & 1.00 & &  & 1.00 & &  & 1.00 & &  & 1.00 &\\
\\
O$^{0}$(6300) & [N II] & $^{*}$8.60 & $\pm$3.10(-6) & [N II] & $^{*}$1.69 & $\pm$0.24(-6) & [N II] & $^{*}$7.31 & $\pm$2.34(-6) & [N II] & $^{*}$5.94 & $\pm$1.82(-6) & [N II] & $^{*}$2.96 & $\pm$0.81(-5)\\
O$^{0}$(6363) & [N II] & $^{*}$8.22 & $\pm$2.98(-6) & [N II] & $^{*}$8.24 & $\pm$4.21(-7) & [N II] & $^{*}$8.23 & $\pm$2.65(-6) & [N II] & $^{*}$5.87 & $\pm$2.46(-6) & [N II] & $^{*}$2.93 & $\pm$0.81(-5)\\
O$^{0}$ & wm & 8.51 & $\pm$3.03(-6) & wm & 1.57 & $\pm$0.23(-6) & wm & 7.56 & $\pm$2.38(-6) & wm & 5.92 & $\pm$1.83(-6) & wm & 2.95 & $\pm$0.79(-5)\\
O$^{+}$(3727) & [N II] & $^{*}$3.79 & $\pm$2.81(-4) & [N II] & $^{*}$1.51 & $\pm$0.49(-5) & [N II] & $^{*}$1.79 & $\pm$1.12(-5) & [N II] & $^{*}$1.54 & $\pm$0.72(-5) & [N II] & $^{*}$6.50 & $\pm$1.73(-5)\\
O$^{+}$(7325) & [N II] & $^{*}$3.67 & $\pm$1.78(-4) & [N II] & $^{*}$1.79 & $\pm$0.71(-5) & [N II] & $^{*}$2.36 & $\pm$1.01(-5) & [N II] & $^{*}$2.24 & $\pm$0.97(-5) & [N II] & $^{*}$4.27 & $\pm$1.94(-5)\\
O$^{+}$ & wm & 3.78 & $\pm$2.65(-4) & wm & 1.65 & $\pm$0.33(-5) & wm & 2.21 & $\pm$0.89(-5) & wm & 2.12 & $\pm$0.85(-5) & wm & 6.25 & $\pm$1.63(-5)\\
O$^{+2}$(5007) & [O III] & $^{*}$1.35 & $\pm$0.71(-5) & [O III] & $^{*}$5.35 & $\pm$1.27(-4) & [O III] & $^{*}$5.78 & $\pm$1.36(-4) & [O III] & $^{*}$3.67 & $\pm$1.46(-4) & [O III] & $^{*}$1.18 & $\pm$0.33(-4)\\
O$^{+2}$(4959) & [O III] & $^{*}$1.32 & $\pm$0.67(-5) & [O III] & $^{*}$5.06 & $\pm$0.96(-4) & [O III] & $^{*}$5.51 & $\pm$1.04(-4) & [O III] & $^{*}$3.54 & $\pm$1.30(-4) & [O III] & $^{*}$1.16 & $\pm$0.26(-4)\\
O$^{+2}$(4363) & [O III] & $^{*}$1.35 & $\pm$0.71(-5) & [O III] & $^{*}$5.35 & $\pm$1.27(-4) & [O III] & $^{*}$5.78 & $\pm$1.36(-4) & [O III] & $^{*}$3.67 & $\pm$1.46(-4) & [O III] & $^{*}$1.18 & $\pm$0.33(-4)\\
O$^{+2}$ & wm & 1.34 & $\pm$0.70(-5) & wm & 5.28 & $\pm$1.16(-4) & wm & 5.71 & $\pm$1.25(-4) & wm & 3.64 & $\pm$1.41(-4) & wm & 1.18 & $\pm$0.30(-4)\\
icf(O) &  & 1.00 & &  & 1.03 &  &  & 1.11 &  &  & 1.17 &  &  & 1.53 & \\
\\
\vdots \\
\\
S$^{+}$ & [N II] & $^{*}$1.47 & $\pm$0.92(-6) & [N II] & $^{*}$1.03 & $\pm$0.34(-7) & [N II] & $^{*}$5.82 & $\pm$3.23(-7) & [N II] & $^{*}$4.71 & $\pm$2.03(-7) & [N II] & $^{*}$2.80 & $\pm$0.67(-6)\\
S$^{+}$(6716) & [N II] & 1.48 & $\pm$0.92(-6) & [N II] & 1.03 & $\pm$0.34(-7) & [N II] & 5.80 & $\pm$3.26(-7) & [N II] & 4.70 & $\pm$2.03(-7) & [N II] & 2.83 & $\pm$0.68(-6)\\
S$^{+}$(6731) & [N II] & 1.47 & $\pm$0.92(-6) & [N II] & 1.03 & $\pm$0.34(-7) & [N II] & 5.84 & $\pm$3.22(-7) & [N II] & 4.72 & $\pm$2.02(-7) & [N II] & 2.78 & $\pm$0.66(-6)\\
S$^{+}$ & [S II] & 1.80 & $\pm$3.01(-6) & [S II] & \nodata &  & [S II] & 1.28 & $\pm$2.04(-6) & [S II] & 5.40 & $\pm$7.72(-7) & [S II] & 4.56 & $\pm$2.77(-6)\\
S$^{+2}$(9532) & [S III] & $^{*}$7.55 & $\pm$2.51(-6) & [S III] & $^{*}$1.77 & $\pm$0.62(-6) & [S III] & $^{*}$6.96 & $\pm$2.27(-6) & [S III] & $^{*}$5.23 & $\pm$1.72(-6) & [S III] & $^{*}$5.20 & $\pm$1.67(-6)\\
S$^{+2}$(6312) & [S III] & $^{*}$7.55 & $\pm$2.51(-6) & [S III] & $^{*}$1.77 & $\pm$0.62(-6) & [S III] & $^{*}$6.96 & $\pm$2.27(-6) & [S III] & $^{*}$5.23 & $\pm$1.72(-6) & [S III] & $^{*}$5.20 & $\pm$1.66(-6)\\
S$^{+2}$ & wm & 7.55 & $\pm$2.51(-6) & wm & 1.77 & $\pm$0.62(-6) & wm & 6.96 & $\pm$2.27(-6) & wm & 5.23 & $\pm$1.72(-6) & wm & 5.20 & $\pm$1.67(-6)\\
icf(S) &  & 1.00 & &  & 1.83 &  &  & 1.81 &  &  & 1.68 &  &  & 1.17 & \\
\enddata
\tablenotetext{a}{wm indicates the calculated mean value for the ionic abundance weighted by the observed flux}
\tablenotetext{b}{an * indicates those values included in the weighted mean}
\end{deluxetable}
\clearpage

\begin{deluxetable}{lr@{}lr@{}lr@{}lr@{}lr@{}lr@{}lr@{}l}
\tabletypesize{\small}
\setlength{\tabcolsep}{0.07in}
\rotate
\tablecolumns{13}
\tablewidth{0in}
\tablenum{5}
\tablecaption{Total Elemental Abundances}
\tablehead{
\colhead{Parameter}\tablenotemark{a} &
\multicolumn{2}{c}{Cn3-1} &
\multicolumn{2}{c}{H2-18} &
\multicolumn{2}{c}{Hb4} &
\multicolumn{2}{c}{Hb6} &
\multicolumn{2}{c}{He2-111} &
\multicolumn{2}{c}{Solar Ref}\tablenotemark{b} &
\multicolumn{2}{c}{Orion Ref}\tablenotemark{c}
}
\startdata
He/H & 5.57 & $\pm$0.70(-2) & 0.120 & $\pm$0.015 & 0.134 & $\pm$0.016 & 0.136 & $\pm$0.017 & 0.221 & $\pm$0.026 & 8.51(-2) & & 9.80(-2) &\\
N/H & 7.75 & $\pm$2.50(-5) & 6.24 & $\pm$1.93(-5) & 5.17 & $\pm$1.36(-4) & 4.09 & $\pm$1.66(-4) & 8.25 & $\pm$2.23(-4) & 6.03(-5) & & 6.03(-5) &\\
N/O & 0.198 & $\pm$0.096 & 0.111 & $\pm$0.025 & 0.787 & $\pm$0.133 & 0.907 & $\pm$0.181 & 2.99 & $\pm$0.70 & 0.132 & & 0.115 &\\
O/H & 3.91 & $\pm$2.66(-4) & 5.62 & $\pm$1.20(-4) & 6.57 & $\pm$1.40(-4) & 4.50 & $\pm$1.65(-4) & 2.76 & $\pm$0.62(-4) & 4.57(-4) & & 5.25(-4) &\\
Ne/H & \nodata &  & 1.12 & $\pm$0.26(-4) & 1.51 & $\pm$0.35(-4) & 7.76 & $\pm$3.25(-5) & 1.27 & $\pm$0.34(-4) & 6.92(-5) & & 7.76(-5) &\\
Ne/O & \nodata &  & 0.199 & $\pm$0.032 & 0.230 & $\pm$0.038 & 0.172 & $\pm$0.030 & 0.461 & $\pm$0.079 & 0.151 & & 0.148 &\\
S/H & 9.02 & $\pm$2.89(-6) & 3.41 & $\pm$1.39(-6) & 1.37 & $\pm$0.52(-5) & 9.60 & $\pm$3.70(-6) & 9.36 & $\pm$2.61(-6) & 1.38(-5) & & 1.48(-5) &\\
S/O & 2.31 & $\pm$1.51(-2) & 6.07 & $\pm$2.43(-3) & 2.08 & $\pm$0.81(-2) & 2.13 & $\pm$0.78(-2) & 3.39 & $\pm$1.05(-2) & 3.02(-2) & & 2.82(-2) &\\
Cl/H & 2.16 & $\pm$0.57(-7) & \nodata &  & 1.30 & $\pm$0.41(-7) & 1.07 & $\pm$0.25(-7) & 1.93 & $\pm$0.58(-7) & 3.16(-7) & & 2.14(-7) &\\
Cl/O & 5.53 & $\pm$3.40(-4) & \nodata &  & 1.97 & $\pm$0.67(-4) & 2.38 & $\pm$0.76(-4) & 7.00 & $\pm$2.22(-4) & 6.92(-4) & & 4.07(-4) &\\
Ar/H & 6.76 & $\pm$2.68(-7) & 1.46 & $\pm$0.29(-6) & 3.91 & $\pm$0.71(-6) & 4.05 & $\pm$1.14(-6) & 4.75 & $\pm$0.85(-6) & 1.51(-6) & & 3.09(-6) &\\
Ar/O & 1.73 & $\pm$1.29(-3) & 2.60 & $\pm$0.48(-3) & 5.96 & $\pm$1.04(-3) & 9.00 & $\pm$1.56(-3) & 1.72 & $\pm$0.28(-2) & 3.31(-3) & & 5.89(-3) &\\
\enddata
\tablenotetext{a}{all abundances are of the form N(X)} 
\tablenotetext{b}{Solar values taken from Asplund et al. 2005.} 
\tablenotetext{c}{Orion values taken from Esteban et al. 1998} 
\end{deluxetable}

\begin{deluxetable}{lcccc}
\tabletypesize{\small}
\setlength{\tabcolsep}{0.07in}
\tablenum{6}
\tablecolumns{5}
\tablewidth{0in}
\tablecaption{Linear Fit Parameters\label{lsf}}
\tablehead{
\colhead{Relation\tablenotemark{1}} & \colhead{Y Intercept} & \colhead{Slope} & \colhead{Correlation Coefficient} & \colhead{Objects}
}

\startdata
\cutinhead{H II Regions $\&$ BCGs}
Ne/H v. O/H & -0.84$\pm$.27 & +1.02$\pm$.03 & +0.96 & 85 \\
S/H v. O/H & -1.63$\pm$.22 & +1.01$\pm$.03 & +0.97 & 73\\
Ar/H v. O/H & -3.03$\pm$.75 & +1.10$\pm$.09 & +0.80 & 82 \\
S/H v. Ne/H & -0.46$\pm$.26 & +0.95$\pm$.04 & +0.96 & 69(2)\\
Ar/H v. Ne/H & -1.26$\pm$.30 & +0.96$\pm$.04 & +0.94 & 74 \\
S/H v. Ar/H & +0.90$\pm$.19 & +0.97$\pm$.03 & +0.96 & 69 \\
Ne/O v. O/H & -0.84$\pm$.27 & +0.017$\pm$.03 & +0.055 & 85\\
Ar/Ne v. Ne/H & -1.26$\pm$.30  & -0.036$\pm$.04  & -0.10 & 74 \\
\cutinhead{Planetary Nebulae}
Ne/H v. O/H & -1.06$\pm$.55 & +1.05$\pm$.06 & +0.81 & 146\\
Cl/H v. O/H & -3.66$\pm$.73 & +1.01$\pm$.08 & +0.72 & 134\\
Ar/H v. O/H & -1.76$\pm$.68 & +0.94$\pm$.08 & +0.70 & 153 \\
Cl/H v. Ne/H & +0.38$\pm$.71 & +0.59$\pm$.09 & +0.51 & 128 \\
Ar/H v. Ne/H & -1.77$\pm$.47 & +0.57$\pm$.06 & +0.63 & 147 \\
Cl/H v. Ar/H & -0.35$\pm$.40 & +0.85$\pm$.06 & +0.77 & 132 \\

\enddata

\tablenotetext{1}{X/H values are expressed as 12+log(X/H), while X/Y values are given as log(X/Y), where X and Y are metals.}
\end{deluxetable}

\begin{deluxetable}{lcc}
\tabletypesize{\small}
\setlength{\tabcolsep}{0.07in}
\tablenum{7}
\tablecolumns{3}
\tablewidth{0in}
\tablecaption{Model Results: Yields from Massive Stars and SNIa\label{models}}
\tablehead{
\colhead{Relation} & \colhead{Predicted Slope} & \colhead{Observed Slope} 
}

\startdata
Ne/H v. O/H & +1.11 & +1.02 \\
S/H v. O/H & +1.20 & +1.01 \\
Ar/H v. O/H & +1.06 & +1.10  \\
S/H v. Ne/H & +1.10 & +0.95 \\
Ar/H v. Ne/H & +0.96 & +0.96 \\
S/H v. Ar/H & +1.13 & +0.97 \\
\enddata

\end{deluxetable}

\clearpage

\clearpage

\end{document}